\documentclass[fleqn,usenatbib]{mnras}

\usepackage{newtxtext,newtxmath}

\usepackage[T1]{fontenc}
\DeclareRobustCommand{\VAN}[3]{#2}
\let\VANthebibliography\thebibliography
\def\thebibliography{\DeclareRobustCommand{\VAN}[3]{##3}\VANthebibliography}

\usepackage{float}
\usepackage{caption}
\usepackage{subcaption}

\usepackage{graphicx}	
\usepackage{amsmath}	
\newcommand{\angstrom}{\textup{\AA}}

\title[Disc evolution in stellar clusters]{The evolution of protoplanetary discs in star formation and feedback simulations}

\author[L. Qiao et al.]{
Lin Qiao,$^{1}$\thanks{E-mail: lin.qiao@qmul.ac.uk}
Thomas J. Haworth,$^{1}$
Andrew D. Sellek$^{2}$
and Ahmad A. Ali$^{3}$
\\
$^{1}$Astronomy Unit, School of Physics and Astronomy, Queen Mary University of London, London E1 4NS, UK\\
$^{2}$Institute of Astronomy, Madingley Rd, Cambridge, CB3 0HA, UK\\
$^{3}$Department of Physics and Astronomy, University of Exeter, Stocker Road, Exeter EX4 4QL, United Kingdom
}

\date{Accepted XXX. Received YYY; in original form ZZZ}

\pubyear{2022}

\begin{document}
\label{firstpage}
\pagerange{\pageref{firstpage}--\pageref{lastpage}}
\maketitle

\begin{abstract}
We couple star cluster formation and feedback simulations of a Carina-like star forming region with 1D disc evolutionary models to study the impact of external photoevaporation on disc populations in massive star forming regions. To investigate the effect of shielding of young stellar objects by star forming material, we track the FUV field history at each star in the cluster with two methods: i) Monte Carlo radiative transfer accounting for the shielding of stars from the FUV by the star forming cloud ii) Geometric dilution of the radiation from other stars which ignores shielding effects. We found that significant shielding only occurs for a small fraction of discs and offers protection from external photoevaporation for < 0.5\,Myr. However, this initial protection can prevent significant early gas/dust mass loss and disc radius reduction due to external photoevaporation. Particularly, shielding for 0.5\,Myr is sufficient for much of the solid reservoir to evolve to larger sizes where it will not be entrained in an external wind. Shielding is therefore potentially significant for terrestrial planet formation in retaining the solid mass budget, but the continued stripping of gas when shielding ends could still impact migration and the gas reservoir for giant planet atmospheres. Our models highlight issues with treating all discs in a cluster with a single characteristic age, since shielded objects are typically only the youngest. Our model predicts that the majority of discs in a 2\,Myr Carina-like environment are subject to strong external photoevaporation.

\end{abstract}

\begin{keywords}
protoplanetary discs -- planets and satellites: formation  -- H\textsc{ii} regions -- ISM: clouds -- circumstellar matter 
\end{keywords}



\section{Introduction}



    With over 4500 planets now confirmed around stars other than the sun\footnote{\url{https://exoplanetarchive.ipac.caltech.edu/}}, understanding the formation and evolution of planetary systems is now a key goal and necessary for interpreting the properties of planets being discovered. There is now overwhelming evidence that planets form from discs of material around young stars \citep[e.g.][]{2015ApJ...808L...3A, 2018A&A...617A..44K, 2018ApJ...869L..41A, 2018ApJ...860L..13P, 2019Natur.574..378T, 2021ApJS..257....1O}. However the majority of observations have been directed towards the brighter and more easily resolved nearby ($<200\,$pc) discs, which in turn have acted as the best laboratories to guide theoretical effort to understand planet formation \citep[e.g.][]{2015ApJ...808L...3A, 2015MNRAS.453L..73D, 2018ApJ...860L..13P, 2020MNRAS.499.2015T, 2021ApJ...921L..34M}. 
    
    However, these nearby systems are members of low mass star forming regions/sparse stellar groups such as Taurus/Lupus. In recent years there has been an increasing acknowledgement that most stars form in much larger clusters where the stellar density is higher than for these nearby systems \citep[e.g.][]{2003ARA&A..41...57L, 2008ApJ...675.1361F, 2010ARA&A..48...47A, 2020MNRAS.491..903W}. The higher stellar density can lead to gravitational encounters that can truncate discs \citep[e.g.][]{1993MNRAS.261..190C, 2005A&A...437..967P, 2014MNRAS.441.2094R, 2015MNRAS.449.1996D, 2016ApJ...828...48V, 2018ApJ...859..150R, 2018MNRAS.475.2314W, 2019MNRAS.483.4114C, 2020MNRAS.491..504C, 2020RSOS....701271P}. In massive stellar clusters, massive stars also form, which emit large amounts of UV photons that photoionise and disperse the star forming cloud \citep{1989ApJ...345..782M, 2012MNRAS.427..625W, 2014MNRAS.442..694D, 2015MNRAS.454.4484G, 2016MNRAS.463.3129G, 2018MNRAS.477.5422A, 2021MNRAS.501.4136A, 2021MNRAS.506.2199G, 2022arXiv220100882G}. This injection of energy into the surroundings is called ``feedback'' and the main focus of prior simulations into this has been the effect on the star formation within clouds. However by dispersing the star forming material, embedded young stellar objects (YSOs) with protoplanetary discs can also become exposed to the strong cluster radiation field. If the disc is sufficiently extended and the radiation field sufficiently strong, this external irradiation heats the disc up and drives loss through a wind in a process called external photoevaporation. 
    
    External photoevaporation has been directly observed to be happening for discs in strong UV radiation environments for some time now, such as in the central 0.1\,pc of the Orion Nebular Cluster (ONC) where the far ultraviolet (FUV) radiation field is in the range $10^5-10^7$\,G$_0$\footnote{G$_0$ represents the Habing unit \citep{1968BAN....19..421H} which is the interstellar radiation density in the solar neighbourhood in the range $912-2400$\AA\,\, and 1G$_0$ = $1.6\times10^{-3}$\,erg\,cm$^{-2}$\,s$^{-1}$. This is subtly different to the \citep{1978ApJS...36..595D} measure of the UV radiation field, with 1 Habing being 1.71 Draines.  }. Here the radiation field is so strong that the circumstellar disc is enshrouded in a cometary wind, with the cusp directed towards the exciting UV source \citep{1994ApJ...436..194O, 1998AJ....116..322H, 1999AJ....118.2350H, 2005AJ....129..382S}. These cometary objects are now referred to as proplyds and the mass loss rates associated with them are expected to significantly reduce the disc mass and lifetime \citep[e.g.][]{1999AJ....118.2350H, 2016arXiv160501773G, 2019MNRAS.490.5678C}. If external photoevaporation were to remove material more quickly than viscous spreading can resupply it, then the disc could also be truncated \citep{2007MNRAS.376.1350C, 2017MNRAS.468.1631R, 2018ApJ...860...77E}. 

    The central ONC is considered to be an extreme environment, and the dynamics of the cluster could mean a large number of young stars enter that extreme environment at some point in the cluster evolution \citep{2019MNRAS.490.5478W}. Nevertheless, evidence is now emerging for external photoevaporation in intermediate UV environments that are thought to be more common \citep{2008ApJ...675.1361F, 2020MNRAS.491..903W} with proplyds discovered in the vicinity of the B star 42 Ori in NGC 1977 \citep{2012ApJ...756..137B, 2016ApJ...826L..15K}. There is also recent evidence for external photoevaporation happening on very early timescales in a stellar cluster \citep[$<0.5$\,Myr][]{2021MNRAS.501.3502H}, competitive even with the earliest evidence for planet formation \citep{2018ApJ...857...18S, 2020Natur.586..228S}.
    
    External photoevaporation in weaker/intermediate UV environments is aided by the fact that only small dust grains are entrained in an external photoevaporative wind. This means that in a disc where grain growth has proceeded the extinction in the wind is low, hence favouring effective external photoevaporation \citep{2016MNRAS.457.3593F}.  External photoevaporation may also take place in extremely weak UV environments for very extended discs, where only modest heating is required to unbind material. This is a possible explanation for the break in the surface density profile of IM Lup \citep{2009A&A...501..269P, 2016ApJ...832..110C, 2017MNRAS.468L.108H} and may also explain the kinematics at the CO surface of HD 163296 \cite{2019Natur.574..378T, 2021ApJS..257...18T}. 
    
    In addition to direct observations of externally photoevaporating discs, evidence for external photoevaporation is emerging from the comparison of statistics of discs in clusters, such as masses \citep{2017AJ....153..240A, 2018ApJ...860...77E}, radii \citep{2018ApJ...860...77E, 2020ApJ...894...74B, 2021ApJ...923..221O} and disc fractions \citep[e.g.][]{2019MNRAS.486.4354R, 2020A&A...640A..27V}. 
    
    To understand these observations we require theoretical models of external photoevaporation. Analytic models were originally developed by \cite{1998ApJ...499..758J}. However, photodissociation region (PDR) microphysics is key to determining the heating and hence mass loss rate in external photoevaporation, which cannot be solved analytically. \cite{2004ApJ...611..360A} performed semi-analytic models using pre-tabulated PDR temperatures, which was also the approach taken by \cite{2016MNRAS.457.3593F} in their work on grain entrainment.  Recently, PDR-hydrodynamics simulations have been developed \citep{2016MNRAS.463.3616H}, however none of the above approaches that include PDR heating enable on-the-fly modelling of disc evolution with external photoevaporation. The production of a publicly available grid of mass loss rates called FRIED \citep{2018MNRAS.481..452H} which can be coupled with disc evolutionary models has alleviated this issue. As a result, FRIED has stimulated work on the relative importance of external photoevaporation and fly-by's \citep[external photoevaporation is generally thought to dominate][]{2001MNRAS.325..449S, 2018MNRAS.478.2700W}, on the evolution of dust within externally photoevaporating discs \citep{2020MNRAS.492.1279S}, on the evolution of discs in different environments \citep{2018MNRAS.475.5460H, 2019MNRAS.490.5678C, 2020MNRAS.491..903W, 2021MNRAS.501.1782C} and even on the impact of reduced disc lifetimes on early stellar evolution \citep{2021MNRAS.508.3710R}. 
    
    The majority of disc evolutionary models with external photoevaporation assume that the external UV radiation field is constant. In reality though the UV radiation field could vary substantially over a disc's lifetime, both due to the fact that star formation is an ongoing process, that stars dynamically evolve in a cluster and that young stars are embedded for some time. \cite{2006ApJ...641..504A} used N-body simulations to constrain the average UV field and impact of tidal encounters in clusters.  \cite{2011PASP..123...14H} also computed elliptical orbits of stars around a strong UV source and the orbit-averaged UV field. This has been built on recently by including external photoevaporation as a function of time in n-body simulations of stars \citep{2019MNRAS.485.4893N, 2019MNRAS.485.1489W, 2019MNRAS.490.5678C, 2021MNRAS.501.1782C}, only with all stars being initialised at the same time and computing the UV field at each star using geometric dilution of the radiation from all UV sources in the cluster. \cite{2021arXiv210107826C} went one step further, simulating the collapse of a giant molecular cloud in an SPH hydrodynamical model that self-consistently formed sink particles representative of stars. They found that ongoing star formation is necessary for massive discs to exist beyond the very early cluster stages if the cluster contains massive stars. However, this simulation did not model the effect of radiation feedback on the gas cloud and again calculated the UV field at each star in the cluster assuming geometric dilution (assuming that the cloud is optically thin).  In all of these calculations the disc itself is not resolved in the star formation model, rather the time varying UV field at each star is computed and then retrospectively coupled with FRIED and a disc evolutionary model. Efforts are being made to do this in dynamical models in the absence of external photoevaporation \citep[e.g.][]{2017ApJ...846....7K, 2018MNRAS.475.5618B, 2021arXiv211005501E}, but the expense of doing direct star/disc formation and PDR driven external photoevaporation is prohibitively computationally expensive at present. 
    
    In this paper we follow the evolution of the UV radiation field incident upon stars in a star cluster formation and feedback simulation and compute the evolution of the discs including external photoevaporation due to that UV field. For the first time, our model includes photoionisation and radiation pressure feedback from the stellar cluster and, because it computes the radiation field at each star in the cluster using Monte Carlo radiative transfer, accounts for shielding of the discs by the star forming cloud. This is also the first time that decoupled dust-gas disc evolution has been included in a simulation of external photoevaporation in clusters. (Although the cluster simulation still assumed a constant dust to gas mass ratio as described in section \ref{feedback_sim}.) We aim to determine how the interplay between stellar feedback, ongoing star formation and shielding influence the role of external photoevaporation of protoplanetary discs. 

\section{Method}

In this paper we couple the results of radiation hydrodynamic simulations of star formation and feedback with circumstellar disc evolutionary models. Here we review our models of star formation and feedback (section \ref{feedback_sim}) and disc evolution (section \ref{disc_model}), and describe the approach used to couple them.

\subsection{Star formation and feedback simulations}
\label{feedback_sim}
For the modelling of giant molecular cloud collapse, star formation and feedback, we use one of the existing simulations from \citet{2021MNRAS.501.4136A} which we briefly summarise here. Their calculations were run using the Monte Carlo radiative transfer (MCRT) and 3D grid based hydrodynamics code \textsc{torus} \citep{2019A&C....27...63H}.  The feedback in the models considers photoionisation \citep[which is generally thought to be the dominant feedback mechanism in H\,\textsc{ii} regions prior to supernovae, e.g.][]{2014MNRAS.442..694D, 2020MNRAS.492..915G, 2021arXiv211005492B} and radiation pressure but not winds or protostellar jets \citep[the latter of which can be important prior to the formation of massive stars][]{2022arXiv220100882G}. Note that the radiation pressure force is calculated using both the contributions from dust and gas (through H/He absorption and electron scattering) and that the dust and gas are dynamically coupled in these models. In \textsc{torus} both the stellar and diffuse radiation fields are included with a polychromatic treatment \citep{2012MNRAS.420..562H}, which as we will discuss below enables the calculation of the FUV radiation field in each cell on the grid to feed into disc evolutionary models with external photoevaporation.  

The \citet{2021MNRAS.501.4136A} star formation and feedback simulation generates sink particles based on the  algorithm in \citet{2015MNRAS.448.3156H} which in turn is based on that of \cite{2010ApJ...713..269F}. In the \citet{2021MNRAS.501.4136A} models each sink particle represents a cluster of stars as the individual stars are not resolved in this model due to the spatial resolution. Stars in any given sink therefore move together, but each star's mass/radius/luminosity evolution is considered independently using MIST evolutionary models \citep{2016ApJ...823..102C}. Each sink has a mass reservoir available to form stars which increases via accretion, enabling further star formation. The mass reservoir is converted into stars with an efficiency of 0.3, motivated by e.g. \cite{2003ARA&A..41...57L}. The star masses generated in each sink are drawn from a pre-tabulated list that is randomly sampled from a \cite{2003PASP..115..763C} initial mass function (IMF), for details see section 2.1 of \citet{2021MNRAS.501.4136A}. A sink is considered as a significant UV source (and hence included in the MCRT) if it starts to contain any stars $>$ 8 \(\textup{M}_\odot\). Once this condition is met, all stars (including the low mass ones) contribute to the radiation in the sink. Sinks containing only low mass stars are not included in the MCRT to reduce the computational expense, but do contribute gravitationally.  

\begin{figure*}
    \centering
    \includegraphics[width=2.\columnwidth]{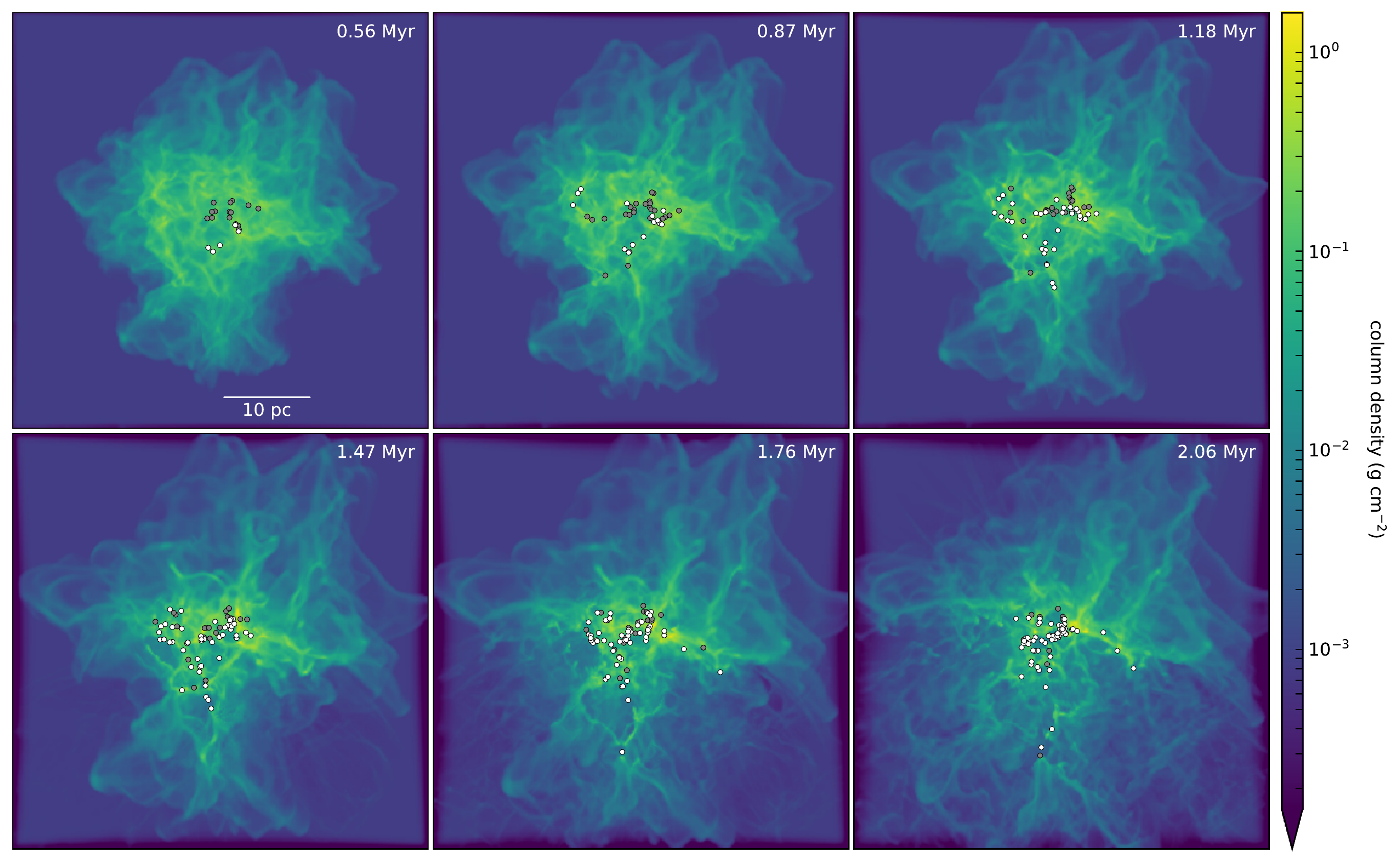}
    
    \includegraphics[width=2.\columnwidth]{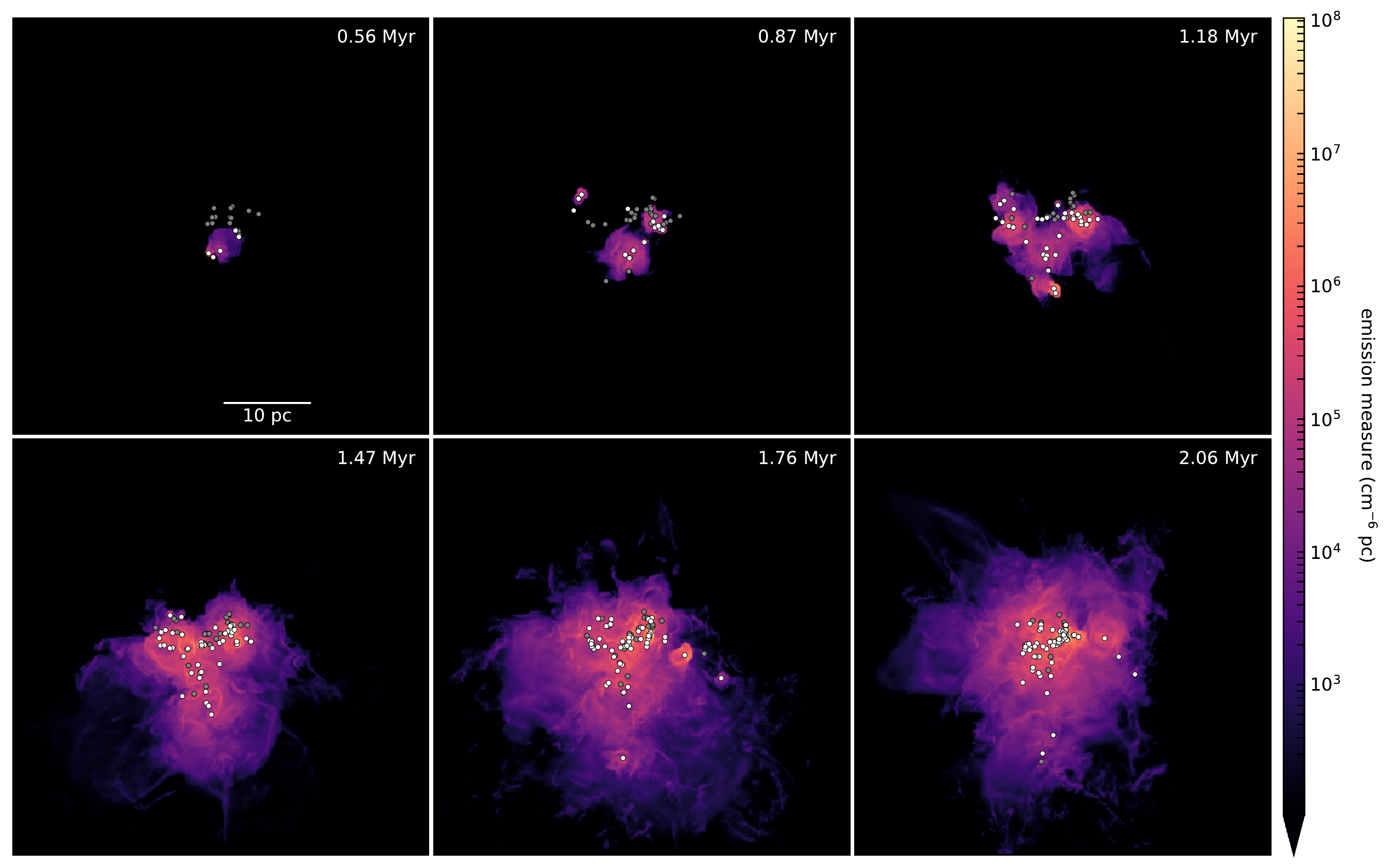} 
    \caption{Snapshots at various cluster ages from the star cluster formation and feedback model from \protect\cite{2021MNRAS.501.4136A}. ($t=0$ is when the first star in the cluster was generated.) The upper set of panels is the column density and the lower set of panels is the integrated electron density squared along the line of sight ($\int n_e^2 dz$), which is a proxy for the free-free continuum emission and highlights where gas has been ionised by feedback. Points represent cluster sinks, with the white points containing ionising/radiating sources. }
    \label{fig:cluster}
\end{figure*}

The initial condition of the \citet{2021MNRAS.501.4136A} model that we utilise was a spherical cloud with mass $M = 10^5$ \(\textup{M}_\odot\), radius $R$ = 11.9 pc, gas and dust temperature of 10 K. A random Gaussian turbulent velocity field was applied following \citep{2002MNRAS.332L..65B}, with the virial parameter $\alpha_{vir} \equiv 2\frac{E_{kin}}{E_{grav}} = 2$ (which means the kinetic energy equals the gravitational energy). This then collapses under gravity, produces stars and is subject to feedback as summarised above. See section 2.2 of \citet{2021MNRAS.501.4136A} for more details of the initial simulation conditions. \citet{2021MNRAS.501.4136A} considered various metallicities from 0.1 to twice solar, however here we focus on the Z / \(\textup{Z}_\odot\) = 1 case, assuming a constant dust-to-gas mass ratio = 0.01. The spatial resolution of the simulation, i.e. the grid cell size is 0.18 pc, and the sink accretion radius is 0.45 pc. Figure \ref{fig:cluster} shows the evolution of the simulated stellar cluster, with the column density in the upper panels and a measure of the free-free emission given by the integral of the square of the electron density along the line of sight through the grid $\int n_e^2 dz$ in the lower panels.

The scale of star formation in this cloud and the resulting stellar population makes it somewhat representative of a massive star forming complex like Carina \citep[e.g.][]{2008hsf2.book.....R}. Furthermore, the estimated ionising photons emitted in Carina is $\sim10^{51}$s$^{-1}$ \citep{2006MNRAS.367..763S} and this model has $1.1\times10^{51}$s$^{-1}$ at the simulation end time of $\sim2$\,Myr since the formation of the first star. Our model also has a roughly constant star formation rate (SFR) of $\sim0.003$ \(\textup{M}_\odot\)yr$^{-1}$ after 1 Myr (discussed further in section \ref{sec:UVtracks}/Figure \ref{fig:star_form_rate}) which is of the same order of magnitude of that in Carina averaged over the last 5\,Myr of $> 0.008$ \(\textup{M}_\odot\)yr$^{-1}$ \citep{2011ApJS..194...14P}. Given the above, in this paper we refer to the model as a Carina-like environment. 

A significant fraction of stars form in massive stellar clusters that include OB stars \citep{2003ARA&A..41...57L}, and this fraction was likely to be even higher at the peak epoch of star formation at redshift $\sim 2$ \citep{2014ARA&A..52..415M}. However disc statistics and the impact of external photoevaporation in Carina are less well constrained than nearby smaller massive star forming regions like Orion. Our choice of star formation model in this paper is therefore motivated in part by the importance of massive star forming complexes, as well as the fact that the simulation was pre-existing and, as we discuss below, can provide detailed estimates of the FUV field incident upon discs. In future work however we will explore the behaviour of discs in clusters in smaller (ONC-like) massive star forming regions.

\subsubsection{Estimating the FUV history of stars in the cluster}
A key goal of this paper is to determine the role of the giant molecular cloud in shielding the discs around young stars from external photoevaporation. To isolate this, the FUV radiation field strength in each grid cell and hence in the vicinity of any given sink is calculated in two ways. The first simply geometrically dilutes the FUV radiation field from all sources to any given point using an inverse square scaling, neglecting cloud shielding \citep[e.g. similar to][]{2021arXiv210107826C}. For the FUV field from a source to a disc inside the same sink, the inverse square law is applied at a distance of half the cell size (0.09 pc), while to a disc in another sink, it is the sink-sink distance. The second uses a Monte Carlo estimator that accounts for the absorption/scattering effects of the cloud. Monte Carlo radiative transfer breaks the energy output of stars into ``photon packets'' which are emitted from sources and undergo a (physical) random walk through the grid until they escape \citep[for more information see][and references therein]{2019A&C....27...63H}. These photon packets have a specific frequency and so we can track those which fall into the FUV range and determine the FUV strength throughout the grid. In this scheme the FUV field in any given cell on the grid is calculated using the following equation \citep{2019MNRAS.487.4890A}: 
\begin{equation}
    FUV(G_{0}) = \frac{1}{H} \int_{912 \, \angstrom}^{2400 \, \angstrom} 4 \pi J_{\lambda} \,d\lambda = \frac{1}{H}\frac{\epsilon}{\Delta t V} \sum_{912 \, \angstrom}^{2400 \, \angstrom} \ell ,
\end{equation}
where $FUV(G_0)$ is the FUV field in the units of the Habing field ($G_0$), $H = 1.63 \times 10^{-3}$ erg s\textsuperscript{-1} cm\textsuperscript{-2}, $J_{\lambda}$ is the mean intensity, $V$ is the cell volume and $\ell$ is the path length travelled between photon events \citep[these events include scattering, absorption and cell-boundary crossing, e.g. ][]{2003MNRAS.340.1136E} by a Monte-Carlo photo packet with energy $\epsilon$ and in time step $\delta t$. In brief, as photon packets in the FUV range traverse a cell their contribution to the energy density and hence mean intensity in that volume of space is recorded and contribute to the total in that region. The FUV field at the location of each sink in the cluster is computed as the FUV field at the grid cell that contains the sink. Note that though the sink positions are continuous, the grid cells are discrete, meaning that one grid cell has the same the FUV field strength anywhere. This means we essentially assume a distance of less than one cell size (< 0.18 pc) between the FUV source and a typical disc within the same sink.

The FUV field is computed in both of the ways described above for each sink particle. This captures processes such as the sinks moving, the gas around them being dispersed and new UV sources ``switching on''. Due to the discrete nature of the grid there can be small scale, but spatially abrupt changes in the FUV incident upon a sink as it crosses a cell boundary, however unless the sink is genuinely crossing from one physical regime to another (e.g. embedded in a cloud to exposed in the H\textsc{ii} region) this is at a level that has negligible impact upon the external photoevaporative mass loss rate. A discussion of the UV tracks themselves is given in section \ref{sec:UVtracks}. The FUV field track at each sink, alongside with the stellar masses formed in the sink is fed into the disc evolutionary models (details see section \ref{disc_model}). The entire cluster is simulated for 2\,Myr, with a total of 84 sinks generated throughout the simulation containing 4444 stars.

\subsection{Disc evolutionary models with external photoevaporation}
\label{disc_model}
We use the model of \cite{2020MNRAS.492.1279S} for the evolution of gas and dust in the disc, including the effect of external photoevaporation. It is based on the gas viscous evolution and dust grain growth and radial drift in the model of \citet{2017MNRAS.469.3994B}, which uses the two-population model in \citet{2012A&A...539A.148B} for treating dust. The code keeps track of the the dust surface density at each time step subject to the mechanisms of radial migration (including radial drift caused by the drag with gas, advection with the viscous flow of the gas, diffusion) and entrainment in the photoevaporative wind. The calculation of gas mass loss via external photoevaporation uses the FRIED grid \citep{2018MNRAS.481..452H}. The FRIED grid provides mass loss rates for externally photoevaporated discs across a range of parameters: disc sizes (1-400\,au), disc masses (or surface densities, for a given disc size, disc masses can be converted to surface densities), stellar masses (0.05 - 1.9\,\(\textup{M}_\odot\)) and FUV field strengths ($10 - 10^{4}$\,G$_0 $). FRIED accounts for the fact that only small grains are entrained in the wind, meaning that when grain growth occurs there is low dust-to-gas mass ratio and dust cross section in the wind (see \citet{2016MNRAS.457.3593F} and Table 1 of \citet{2018MNRAS.481..452H}). For each disc simulated, at each time step, we determine the mass loss rate by linearly interpolating the FRIED grid in three dimensions: disc size $R_{D}$, outer disc surface density $\Sigma_{\textrm{out}}$ and the FUV field strength provided by the FUV track of the specific host star from \citet{2021MNRAS.501.4136A}. For the detailed interpolation scheme in disc size and surface density, see section 2.2.1 of \citet{2020MNRAS.492.1279S}. The original code in \citet{2020MNRAS.492.1279S} only interpolates FRIED in these two dimensions with constant FUV field throughout the disc evolution. To calculate the mass loss rate with time-varying FUV field, we modified the code to also interpolate in the FUV field dimension. We do not interpolate in the dimension of stellar mass because 4-D interpolation makes the code run significantly slower. Instead, for each disc with its host star mass $M_{*}$, we chose the closest value of stellar mass to $M_{*}$ on the FRIED grid (note that the stellar mass does not  evolve over the duration of the simulation to a degree that has a significant impact on external photoevaporation). For dust mass loss via external photoevaporation, grains small enough to be entrained in the wind via drag are removed, for a detailed discussion of the approach to dust mass loss via the entrainment in the photoevaporative wind, see section 2.2.3 of \citet{2020MNRAS.492.1279S}. Moreover, because this model keeps full track of the evolution of the dust surface density profile, this calculation of dust mass removed by the wind uses the real time dust-to-mass ratio at each grid point of the disc, instead of assuming a constant value at all time throughout the disc \citep[see equation 13 of][]{2020MNRAS.492.1279S}. 

\subsubsection{Initial conditions}
\label{disc_initial_cond}
For the initial disc surface density profile $\Sigma$ we use the similarity solution of \citet{1974MNRAS.168..603L}:
\begin{equation}
    \Sigma = \Sigma_{0}\left(\frac{R}{R_C}\right)^{-1}\exp{\left(-\frac{R}{R_C}\right)},
\end{equation}
where $\Sigma_0$ is the normalisation constant set by the total disc mass, and $R_C$ is the scale radius, which sets the initial disc size. $R_C$ is a free parameter which we vary with three values: 10\,au, 40\,au, and 100\,au for modelling compact, fiducial and extended discs \citep[e.g.][]{2020A&A...640A...5T, 2021ApJ...917L..10L}. Our model uses a radial temperature profile $T(R)$:
\begin{equation}
    T = T_{0}\left(\frac{R}{R_0}\right)^{-0.5}, 
\end{equation}
where $R_0 = 1$\,au and $T_{0}$ is the temperature at $R_0$ (1\,au), which is set by a prescribed disc aspect ratio $h_{0}$ at 1\,au:
\begin{equation}
    T_{0} = \frac{\mu}{\mathcal{R} h_0}\frac{G M_{*}}{R_0},  
\end{equation}
where we have assumed $h_0 = 0.033$ and a mean molecular weight $\mu=2.5$, $\mathcal{R}$ is the gas constant and $M_{*}$ is the disc's host star mass. We assume a constant $\alpha$ viscosity prescription \citep{1973A&A....24..337S} with $\alpha = 10^{-3}$ \citep[though the value of the viscosity can affect the lifetime of externally photoevaporating discs][]{2020MNRAS.497L..40W}. We set the initial disc mass as $M_{\rm disc} = 0.1$\,M$_{*}$, and the initial dust-to-mass ratio as 0.01. 

With the above initial conditions we evolve discs around every star with mass $M_{*} \leq 1.9$\,\(\textup{M}_\odot\) (the upper mass limit of the FRIED grid) in the star formation and feedback calculation described in \ref{feedback_sim}. For each star in the simulation we run models with each of the three initial $R_C$ and for FUV tracks derived from full the Monte Carlo radiative transfer (which accounts for absorption/scattering by the cloud and the diffuse field) and through simple geometric dilution (which ignores the cloud and diffuse field, see section \ref{feedback_sim}). This gives a total number of models of 6 times the number of stars (3081) in the star formation simulation that are $\leq 1.9$ \(\textup{M}_\odot\).
All discs in the same simulation set have a uniform initial disc size $R_C$ and the initial disc mass scales with the stellar mass, so more massive stars have more gravitationally bound discs.

Each sink particle can contain multiple stars and add more as it accretes material. As soon as a new star is added to the sink particle its disc begins evolving with the specific initial conditions of the model.  At each time step, the FUV field strength incident upon the disc is obtained by linearly interpolating the FUV field track of the corresponding sink position (i.e. once formed all stars in the same sink particle are exposed to the same UV field). An upper and a lower cap on the incident UV field of $10^4 \, G_{0}$ and $10  \, G_0$ are imposed due to the limits of FRIED grid. The upper cap of the UV field, although reached quite often in the simulation, is not considered problematic for a few reasons: Firstly the upper cap means we are being conservative in estimating the mass loss rate $\dot{M}$ due to photo-evaporation. Secondly the discs exposed to high UV environment get truncated via photo-evaporation fast enough that qualitatively, the disc evolution would be the same with capped value compared to if the effects of higher FUV values are taken into account. Thirdly, in the FRIED grid the mass loss rates $\dot{M}$ for FUV = $10^{4} \, G_{0}$ are similar to $\dot{M}$ for ONC proplyds (exposed to UV radiation of $\sim 10^5 - 10^6 \, G_{0}$), so we expect a weak dependence on FUV field strength in the high value range. Because the FUV history of entire stellar cluster is calculated by \citep{2021MNRAS.501.4136A} for only $\sim$ 2\,Myr, many discs in the cluster are evolved for less than 2\,Myr, with the ones around stars born later in the cluster evolving for much less than 2\,Myr in our simulation. However in high UV environments discs evolve on very short timescales \citep[e.g.][]{2020A&A...640A..27V, 2021MNRAS.501.3502H} and many key regions of feedback and star formation such as the ONC are $<2\,$Myr, making the duration of the star formation simulation sufficient for this first investigation. Furthermore the main stellar clusters in Carina Tr 14 and Tr 16 both appear to be dynamically young \citep{2019MNRAS.486.4354R}. However, running these star formation and feedback models for longer is an issue that will be addressed in subsequent work. 

Table \ref{sum_sim} summarises all the subsets of disc simulation carried out.

\begin{table}
 \centering
 \begin{tabular}{|c|c|c|} \hline \hline
 \textbf{Simulation Set} & \textbf{Initial $R_C$} & \textbf{FUV track} \\ \hline \hline
 a & 10\,au & MCRT \\ 
 b & 10\,au  & GD \\
 c & 40\,au  & MCRT\\ 
 d & 40\,au & GD\\ 
 e & 100\,au & MCRT\\ 
 f & 100\,au & GD\\ \hline \hline

  \end{tabular}
  \caption{Summary of all simulation sub-sets, MCRT stands for Monte Carlo Radiative Transfer, GD stands for Geometric Dilution}
 \label{sum_sim}
\end{table}

\subsection{Calculating observers' equivalents of disc properties}
\label{observer}
The size and mass inferred for a protoplanetary disc is observation specific, varying for different gas lines and continuum wavelengths due to processes such as the disc chemistry, grain growth and radial drift \citep[e.g.][]{2011A&A...529A.105G, 2016A&A...588A..53T, 2019A&A...629A..79T}. We therefore cannot directly compare our known model disc masses and radii with observations. Many recent disc surveys have been in the (sub-)millimetre continuum with ALMA \citep[e.g.][]{2016ApJ...828...46A, 2017AJ....153..240A, 2018ApJ...860...77E, 2020AJ....160..248A, 2020A&A...640A..27V, 2021ApJ...923..221O} and so it is beneficial to compare our models in a way that is consistent with those. To this end we follow the approach of \cite{2005ApJ...625..414T}, \cite{2016A&A...588A..53T} and \cite{2020MNRAS.498.2845S}, wherein the disc is assumed to be face-on with total flux from the inner disc to some outer radius as 
\begin{equation}
    F_\nu = \frac{1}{d^2}\int_{R_{\textrm{in}}}^{R_{\textrm{out}}}2\pi R  B_\nu(T)(1 - e^{-\tau_{\nu}})  \textrm{d}R. 
    \label{equn:fluxIntegral}
\end{equation}
Here $B_\nu(T)$ is the Planck function at disc temperature $T(R)$, $\tau_\nu = \kappa_{\nu}\Sigma_\textrm{dust}$ is the optical depth, where $\kappa_{\nu}$ is the continuum opacity at the frequency of observation, $d$ is the distance of the observer. $\kappa_{\nu}(a)$ in equation \ref{equn:fluxIntegral} is as used by \citet{2019MNRAS.486L..63R, 2019MNRAS.486.4829R}, which were calculated for the maximum grain size $a_\textrm{max}(R)$ for compact grains with composition based loosely on \citet{1994ApJ...421..615P} and follows the same methodology and optical constants as \citet{2016A&A...588A..53T}. We consider a wavelength of 850\,$\mu$m in these continuum flux estimates. After the flux density $F_{\nu}$ is computed, it is then converted to a dust mass following \cite{1983QJRAS..24..267H} for optically thin ($\tau_{\nu}<<1$) dust emission:
\begin{equation}
\label{equn:M_dobs}
    M_{\textrm{d,obs}} = \frac{F_\nu d^2}{B_\nu(T_{\textrm{dust}})\kappa_\nu}
    ,
\end{equation}
which is the approach used for mass estimates in recent ALMA surveys, where the typical dust temperature is assumed to be 20\,K \citep[e.g.][]{2016ApJ...828...46A, 2018ApJ...860...77E}, and a constant opacity $\kappa_{\nu} = 0.353$\,m$^2$\,kg$^{-1}$ is used in equation \ref{equn:M_dobs} \citep{1990AJ.....99..924B}. However \cite{2021MNRAS.503.4172H} demonstrated that for proplyds (discs in strong UV environments) their enhanced heating due to the external radiation field means that this 20K assumption can lead to substantial overestimates of the disc mass (e.g. within 0.1\,pc of $\theta^1$C). Nevertheless we employ an assumed dust temperature of 20\,K to be consistent with common observational practice.

For observed disc radii we again follow observational practice and define the observed disc radius as that at which the flux profile (equation \ref{equn:fluxIntegral}) encompasses 68\,per cent of the total disc flux, which we denote as $R_{68}$ \citep[e.g.][]{2016A&A...588A..53T}. 

\begin{figure*}
    \centering
    \includegraphics[width=2\columnwidth]{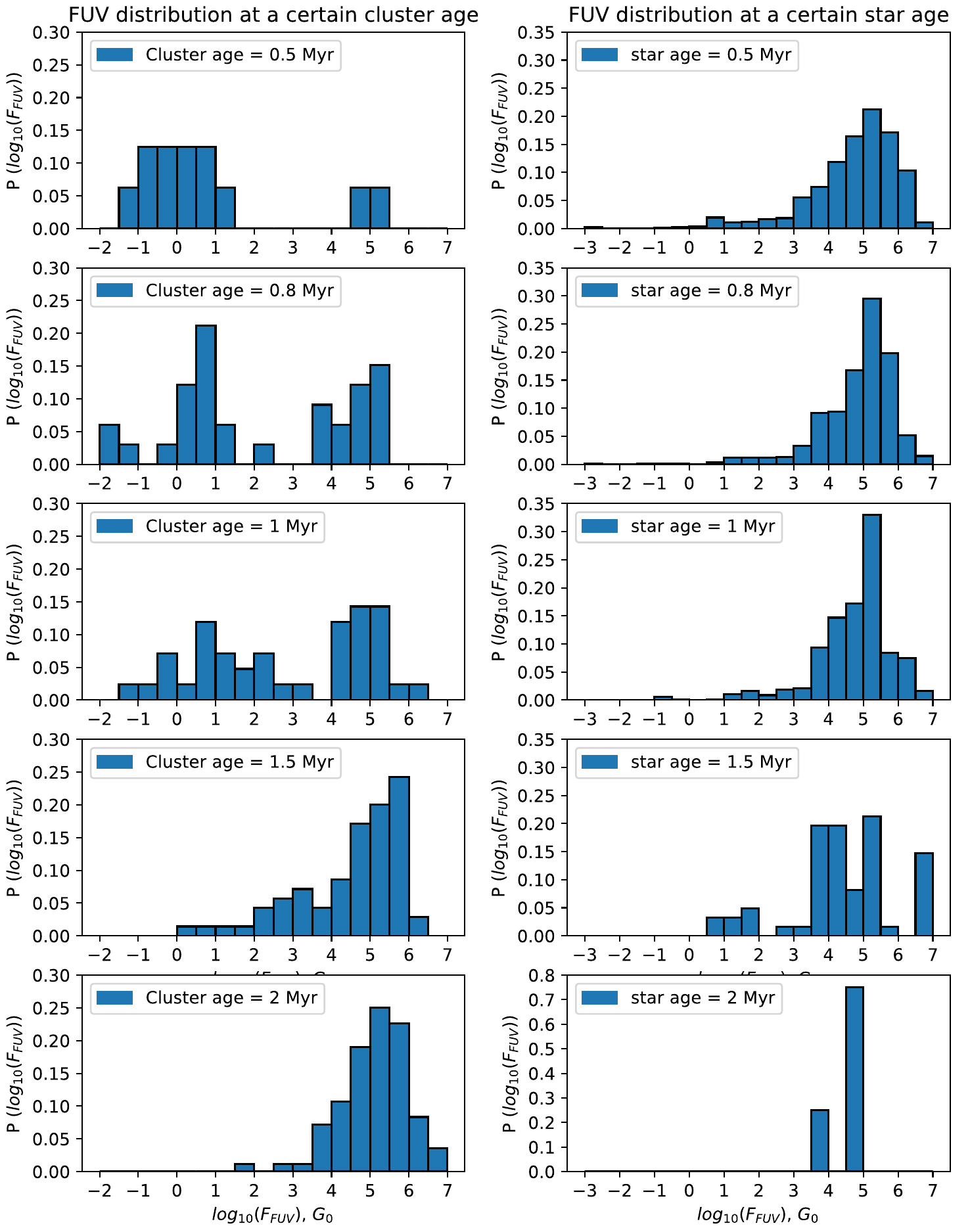}
    \caption{Left: Probability distribution of the FUV field strength at all simulated stars in the cluster at the cluster age of 0.5\,Myr (top), 1\,Myr (middle) and 1.5\,Myr (bottom). Right: Probability distributions at all stars with star age 0.5\,Myr (top), 1\,Myr (middle) and 1.5\,Myr (bottom). The distributions include all stars in the cluster simulation, regardless of whether they have mass $<$1.9\,\(\textup{M}_\odot\)}
    \label{fig:FUV_stats}
\end{figure*}

\section{Distribution of UV fields irradiating the simulated stellar population}
\label{sec:UVtracks}
In this section we provide an overview of the FUV radiation tracks for all stars in the star formation and feedback simulation. These tracks feed into the disc evolution and external photoevaporation models for each star, so it is useful to summarise them before discussing the discs themselves. Note that for brevity we focus here on the FUV tracks calculated using the Monte Carlo estimator (which accounts for the \textit{shielding effects} of the molecular cloud) rather than the simple geometric dilution tracks. Here we include all stars simulated, but only ones with mass < 1.9 \(\textup{M}_\odot\) will have discs evolved around them, as mentioned in section \ref{disc_initial_cond}.

The left panels in figure \ref{fig:FUV_stats} show the probability distribution of FUV field strengths incident upon stars (and hence discs) at  different stellar cluster ages (i.e. at 0.5, 0.8, 1, 1.5 and 2\,Myr since the formation of the first star from top to bottom). At 0.5\,Myr (upper left panel) two populations of stars correspond to two types of FUV environments at this early cluster age. The left side population of stars are in low FUV environments, mostly because they are embedded in gas and hence shielded, whereas even at this early time the right side population indicates a fraction of stars exposed to strong FUV radiation fields due to the feedback driven dispersal of the cloud \citep[such early external photoevaporation is observed towards NGC 2024,][]{2020A&A...640A..27V, 2021MNRAS.501.3502H}. As the cluster age increases, with further gas dispersal and more radiation sources formed, more stars end up in locations with higher FUV values, hence the two populations gradually merge into a single one peaked around high FUV values (as indicated by the distributions at cluster age of 1.5 and 2\,Myr shown in the fourth and firth rows in the left panels of Figure \ref{fig:FUV_stats}). At 2\,Myr, most stars (even newly generated ones) are exposed to high FUV radiation.

\begin{figure}
    \centering
	\includegraphics[width=\columnwidth]{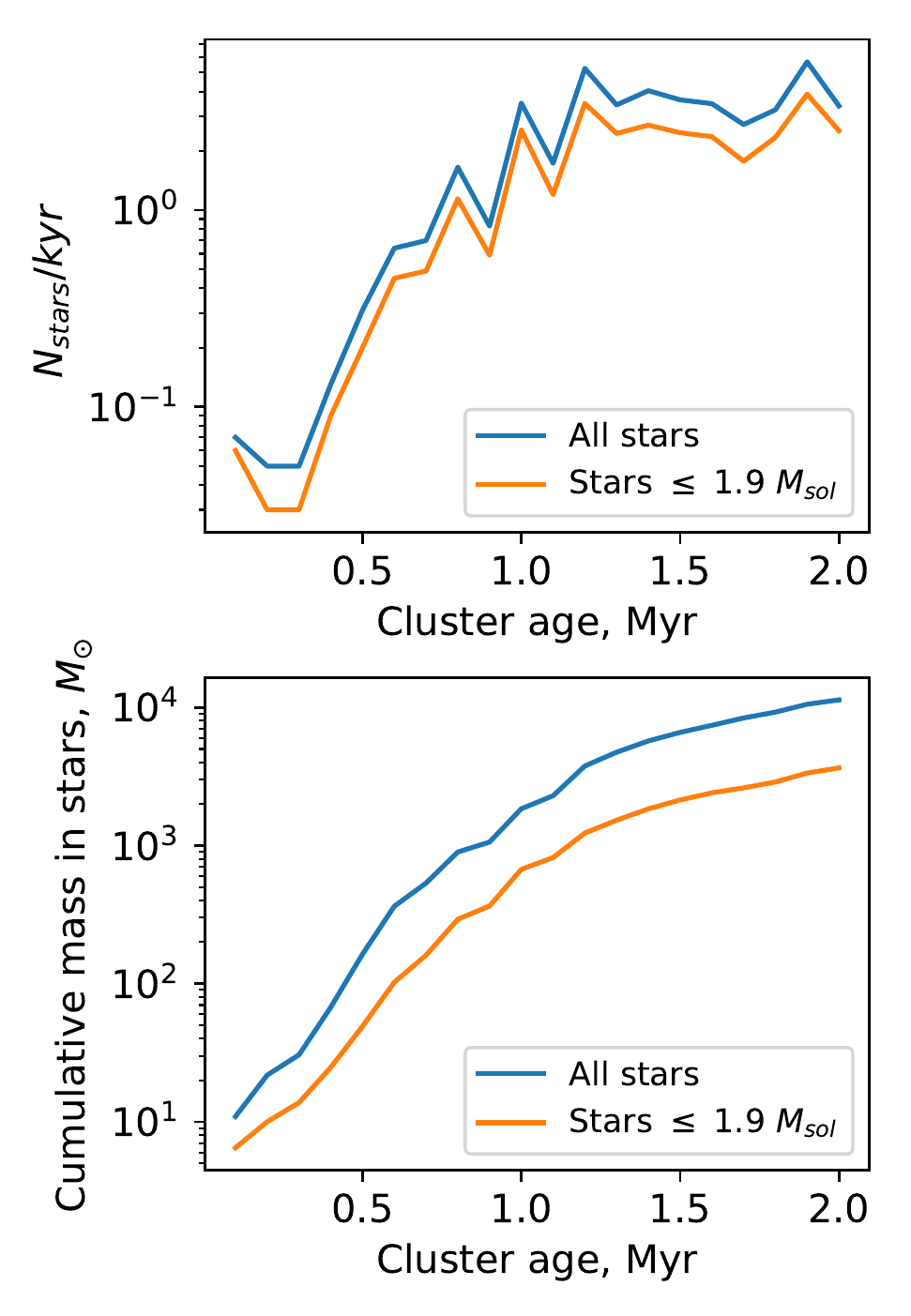}
    \caption{Top: star formation rate in the unit of number of stars/yr as a function of cluster age. Bottom: cumulative star formation in mass as a function of cluster age. The blue line is all stars and the orange line only stars with mass $\leq1.9$\,M$_\odot$, which is the upper limit on stellar mass considered in our disc evolutionary models.} 
    \label{fig:star_form_rate}
\end{figure}

The right panels in figure \ref{fig:FUV_stats} show the probability distributions of FUV fields incident upon star-discs as a function of the stellar age (i.e. the FUV field all stars are exposed to at a given time since their formation, as opposed to at the time since the formation of the first star in the left panels). At 0.5\,Myr (top right panel) the distribution is not split into two populations as in the plot of cluster age = 0.5\,Myr (top left panel). The distribution is peaked around a high FUV value with a small fraction of stars still exposed to low FUV indicated by the left tail. This implies that by a star age of 0.5\,Myr most stars are exposed to strong UV radiation fields in a massive star forming complex like Carina. Embedding any given star therefore only offers protection from external photoevaporation for a short time. 

One key point from Figure \ref{fig:FUV_stats} is the difference between the distribution for a given cluster age (which is what is generally assigned to all stars in a cluster) and distribution at a given star age. The embedded stellar component of a cluster is $<0.5$\,Myr (even in a 1.5\,Myr cluster) so only recently formed. Discs in Carina like environments are therefore expected to be irradiated on a timescale in competition with even the earliest evidence for planet formation in discs \citep{2018ApJ...857...18S, 2020Natur.586..228S}.

Figure \ref{fig:star_form_rate} shows the rate of star formation (in number of stars/yr, upper panel) and cumulative mass in stars (lower panel) as a function of cluster age. The star formation rate increases quickly for the first 1\,Myr to a total mass in stars of $\sim2\times10^3$\,M$_\odot$ before levelling off at a star formation rate of around 3 stars per kyr.

\section{Results}
Here we discuss the nature of protoplanetary disc evolution in our simulated stellar cluster. We begin by looking at the detailed evolution of some representative discs in \ref{sec:detailed} and follow this with an analysis of how the cluster properties evolve with time in \ref{sec:discStats}. 

\subsection{Overview of disc evolution for different types of UV track}
\label{sec:detailed}
\begin{figure*}
    \vspace{-0.4cm}
    \hspace{-0.6cm}
    \includegraphics[width=2\columnwidth]{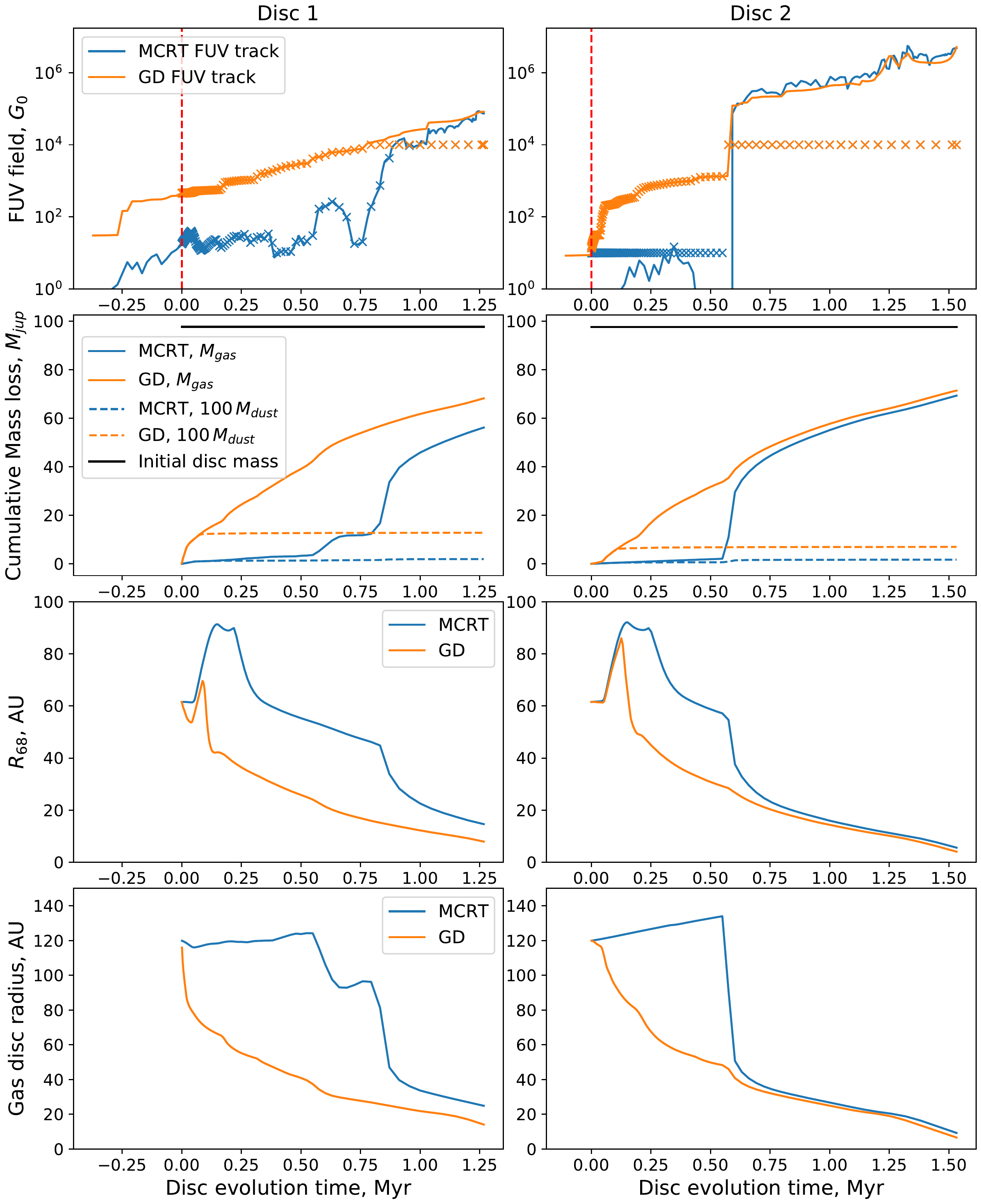}
    \vspace{-0.2cm}
    \caption{The evolution of two selected discs with i) shielding of the cloud included since the FUV track is computed with Monte Carlo radiatve transfer (MCRT) (blue) and ii) no shielding since the FUV track is computed using Geometric Dilution (GD) from each UV source (orange). Negative time corresponds to the FUV history in the location of the disc's sink before the specific star in the cluster sink is generated. First row: the solid lines indicates the original the FUV radiation fields at the disc location provided by the star formation and feedback simulations of \citet{2021MNRAS.501.4136A}, and the crosses indicates the interpolated values used at each disc evolution simulation time step. Second row: cumulative mass loss in gas (solid curves) and 100 $\times$ cumulative mass loss in dust due to external photoevaporation only (dashed curves). Third row: Observer's equivalent of disc size, $R_{68}$, defined in section \ref{observer}. Fourth row: modeled physical gas radius, defined as the radius at which the disc contains 95\, per cent of the total gas mass.}
    \label{fig:selectec_discs}
\end{figure*}

In this section we provide a detailed view of the evolution of two specific discs from the cluster. These will be referred to as disc 1 and disc 2 and represent two characteristic types of FUV histories. Disc 1 starts embedded, with a gradually increasing external radiation field, whereas disc 2 undergoes a much sharper increase in ambient UV field, as illustrated in the upper left and upper right panels of figure \ref{fig:selectec_discs} (note that no discs remain embedded for $>$Myr timescales in our simulated massive stellar cluster, since feedback is effectively dispersing the star forming material). Both discs have similar initial masses (97.65 $M_{\textrm{jup}}$ for disc 1 and 97.62 $M_{\textrm{jup}}$ for disc 2, with the disc to star ratio = 0.1) and initial scaling radius of $R_C$ = 40\,au. For each disc we show the evolution of the FUV track computed using both the MCRT (Monte Carlo radiative transfer, plotted in blue) that accounts for shielding by the molecular cloud and also assuming simple geometric dilution (GD, plotted in orange).  

Disc 1 is exposed to a low ($\sim10$G$_0$) initial FUV radiation environment that gradually increases to $>10^4$G$_0$ over the course of around 0.8\,Myr. Conversely disc 2 is deeply embedded in a $<10$G$_0$ environment for around 0.55\,Myr before being rapidly exposed to a very strong FUV field. Comparing the MCRT and GD tracks allows us to determine the impact of cloud shielding in each scenario. Without shielding, we see that the discs are rapidly in UV environments orders of magnitude stronger. 

The effects of the time varying FUV radiation field represented by these tracks on the discs themselves are shown in the other panels of Figure \ref{fig:selectec_discs}. The second row shows the cumulative mass loss via external photoevaporation of gas (solid lines) and dust (dashed lines, multiplied by a factor $100$ for easier visualisation).  The gas mass loss for both discs evolved with GD tracks (orange) is increasing at a consistently high rate. Conversely for discs with MCRT FUV tracks the cumulative mass loss through external photoevaporation initially stays low due to the cloud shielding, before catching up to high values later as the discs become un-embedded and hence exposed to increasing FUV radiation. This occurs because, although the discs experience similar FUV fluxes to the GD case once they become exposed (at $\sim$0.8\,Myr for disc 1 and $\sim$0.55\,Myr for disc 2), they had remained larger up until those points and had thus retained a reservoir of more easily unbound material which starts to evaporate rapidly, leading to a steep rise in the cumulative mass loss.
The effect of shielding is therefore not to prevent external photoevaporation of the gas disc, however it does delay it, which could be important given the growing evidence for early planet formation in discs \citep[e.g.][]{2018ApJ...857...18S, 2020Natur.586..228S}. Just before the discs with MCRT tracks become un-embedded, disc 1 and disc 2 lost $\sim$ 43 $M_{\textrm{jup}}$ and $\sim$ 30 $M_{\textrm{jup}}$ less gas respectively (which are $>$ 30\,per cent of the initial disc masses), compared to the discs with GD tracks where shielding is not accounted for. This highlights the potentially significant effect of gas shielding in early disc evolution stage.

Dust can only be entrained in the external photoevaporative wind in the very early disc evolution (in the first 0.1 Myr), before the dust particles at the edge of the disc grow too large to be carried away by the wind \citep{2016MNRAS.457.3593F}. The overall dust mass loss due to external photoevaporation is therefore only sensitive to the FUV field in the very early evolution of the disc and so the early shielding has a big impact. The overall dust mass lost due to external photoevaporation for disc 1 is $\sim6$ and $\sim$ 40 $M_{\textrm{earth}}$ for the shielded and unshielded cases respectively (a factor 6.7 difference) and  for disc 2 is $\sim$5 and $\sim$ 22 $M_{\textrm{earth}}$ for the shielded and unshielded cases respectively (a factor 4.4 difference). Despite being shielded for longer, disc 1 is still in a marginally stronger UV field at early times, which is the reason that it loses more dust mass. Therefore it is not just the duration of shielding that is important for preventing dust being stripped in external photoevaporation, but also the magnitude of that shielding. 

Note that although the dust mass loss due to photoevaporative winds stops early on, and can be almost entirely prevented by shielding, a significant amount of dust in the disc continues to be lost via radial drift onto the central star in our models. This is a well known phenomenon in dust evolutionary models such as the \cite{2012A&A...539A.148B} model used here and pressure bumps in the disc are thought to be important for explaining long lived continuum discs \citep[e.g.][]{2020ApJ...888L...4T}. However even in the absence of pressure bumps this rapid disc evolution still gives a reasonable match to observed disc dust gas mass--accretion rates, at least in non-photoevaporating regions \citep{2020MNRAS.498.2845S}. 

The third row of Figure \ref{fig:selectec_discs} shows the observer's equivalent of disc size, $R_{68}$, defined as the radius at which the 850\,$\mu$m flux profile encompasses 68\,per cent of the total disc flux (see section \ref{observer}). Again, shielding from external photoevaporation influences the disc properties, resulting in discs that appear to be more extended early on. Just before being un-embedded and exposed to the strong UV field, $R_{68}$ is around 30\,au larger in the models that include shielding (MCRT, blue) than the ones without (GD, orange) for both disc 1 (at 0.8\,Myr) and disc 2 (at 0.55\,Myr). Note that for all models $R_{68}$ undergoes an initial and temporary apparent growth, which is not actually a reflection of the true disc radius. This feature arises because early on the inner disc loses dust particles through radial drift faster than the outer disc, making the 850\,$\mu$m flux (computed in equation \ref{equn:fluxIntegral}) from the inner disc weaker  and flattening the intensity profile. Therfore a larger radius is needed to encompass 68\,per cent of the total flux. Shortly after radial drift is operating in the outer disc, $R_{68}$ decreases again.

Finally the fourth row shows the gas disc radius, defined as the the radius which contains 95\,per cent of the disc's total gas mass. Here the effects of FUV field and gas shielding is more dramatic than in the case of the dust. For the two discs evolved with MCRT FUV tracks, the gas radii in the embedded low FUV stage is near constant (and even viscously expand for a bit) compared to the sharp initial decrease the GD track cases that have no shielding. Just prior to the un-embedding of the MCRT discs, their gas radii still remained  $\sim$4 times (disc 1) and $\sim$3 (disc 2) times larger than the radii for discs with GD tracks. The MCRT disc radii then decrease steeply with the sudden surge of FUV radiation, showing that the physical gas radii of discs (before getting stripped to tightly bound ones) are strongly sensitive to the FUV radiation environment. 

Overall, these two scenarios highlight that shielding effects are important during the disc's early stage of evolution, providing larger disc masses and radii in that time, which could be important if planet formation happens early. Once the shielding ceases, the discs tend to a similar evolution (of radii, masses, which look similar after 1\,Myr irrespective of shielding) as the unshielded cases, with the exception that almost no dust was depleted in the external photoevaporative wind. In addition to these kind of FUV tracks, our simulations contain a number of stars that are almost instantly exposed to a strong FUV radiation field (particularly those forming at later times), in which case the MCRT and GD tracks are very similar. In those cases, discs would rapidly become proplyds with high mass loss rates and quickly get stripped to small disc size. The continual introduction of new discs into the high UV region is thought to be a key element to resolving the proplyd lifetime problem \citep[we should not observe them if they are so rapidly destroyed][]{1999AJ....118.2350H, 2019MNRAS.490.5478W} and is thought to be happening in the case of proplyds the vicinity of IRS 2b in NGC 2024 \citep{2021MNRAS.501.3502H}

\subsection{Statistics of discs in a simulated cluster}
\label{sec:discStats}
\begin{figure*}
    \vspace{-0.3cm}
    \hspace{-0.6cm}
    \includegraphics[width=2.1\columnwidth]{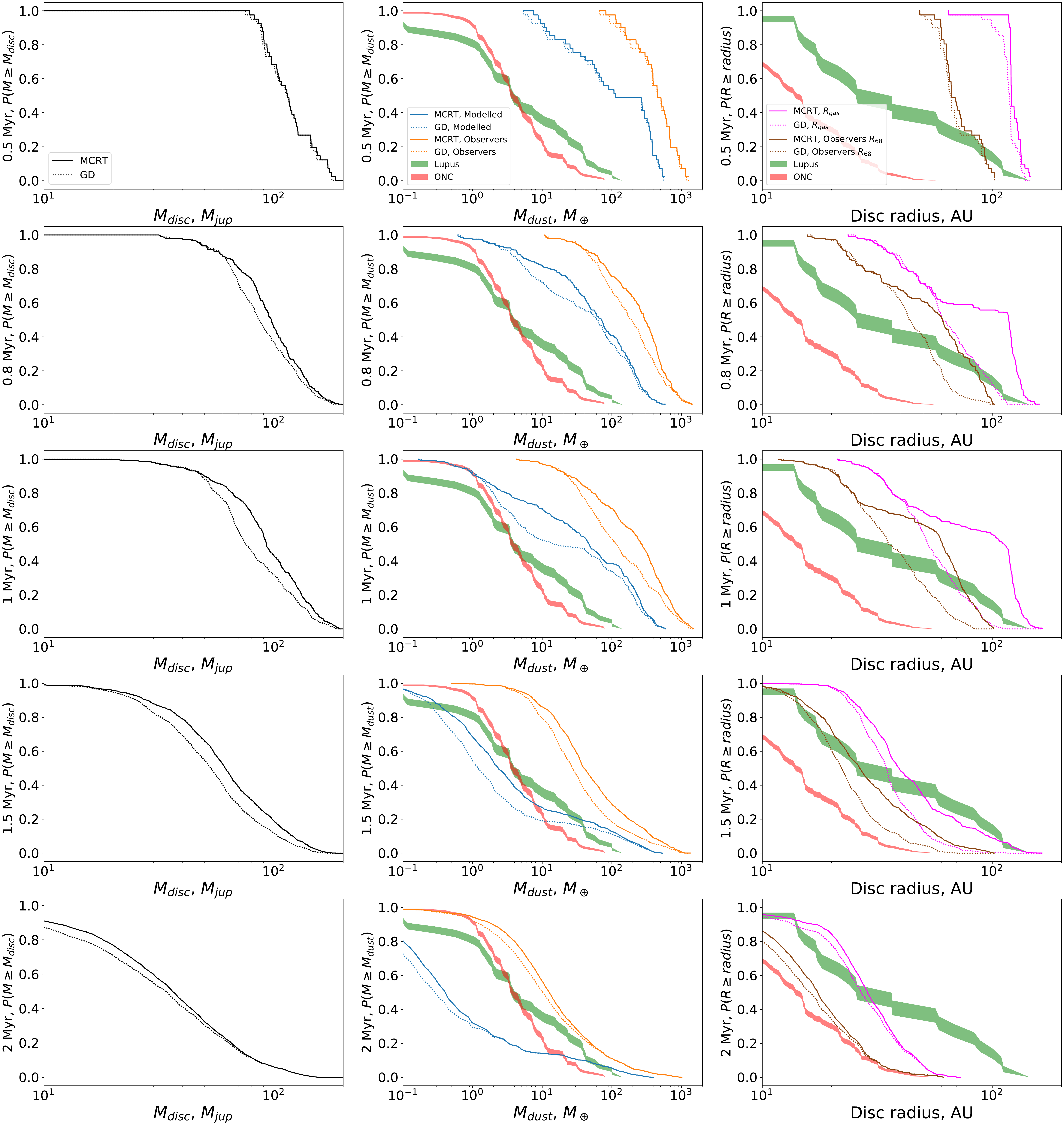}
    \vspace{-0.1cm}
    \caption{The inverse cumulative distribution functions (cdf) of various disc properties for all discs simulated with intial $R_C$ = 40\,au at 5 cluster ages : 0.5\,Myr (first row), 0.8\,Myr (second row), 1\,Myr (third row), 1.5\,Myr (forth row) and 2\,Myr (fifth row). The solid curves show the simulated data set with the MCRT tracks and the dotted curves show data set with GD tracks. The first column shows the cdf of the modelled disc mass. The second column shows the modelled dust mass (blue curves), the calculated observers' equivalent dust mass (orange curves), the green and red curves show the observed dust mass distributions for discs in Lupus and ONC. The third column shows the modelled disc gas radius $R_{\rm gas}$ (magenta curves) and observer's equivalent disc radius $R_{68}$ (brown curves), the green and red curves show the observed disc size distributions for discs in Lupus and ONC.}
    \label{fig:cdf_plots}
\end{figure*}

The previous section focused on analysing the evolution of two specific discs throughout their lifetimes. In this section we look at the statistics of all discs simulated in the entire cluster at various cluster ages (0.5, 0.8, 1, 1.5 and 2\,Myr). We compare the disc properties when evolved with and without shielding (MCRT/solid and GD/dotted respectively) and also compare with observed distributions in Lupus \citep{2016ApJ...828...46A} and the ONC \citep{2018ApJ...860...77E}. The distributions of disc masses, dust masses, and gas and dust disc radii are given in Figure \ref{fig:cdf_plots} for the $R_c=40$\,au set of disc models.

Specifically, the left hand panels in figure \ref{fig:cdf_plots} shows the inverse cumulative distribution function (CDF) of the total disc mass, with the cluster age increasing from top to bottom. The central panels are the inverse CDF of the dust mass; both the true dust mass within the simulation $M_d$ (blue curves) and that which an observer would infer for a face on disc $M_{d, \textrm{obs}}$ (orange curves, see equation \ref{equn:M_dobs}). The right hand panels are the inverse CDF of disc radius in gas (magenta curves) and dust (brown curves). For all variables there is a similar overall behaviour in the difference between the shielded and un-shielded models as a function of time. Up until around 0.5\,Myr the UV radiation in the region is sufficiently weak that the shielded and un-shielded populations are essentially identical. From 0.5 up until around 1\,Myr there is a period where there are strong UV sources, but the star forming cloud is not yet dispersed. In this phase, shielding has the maximum impact and so there is the biggest difference between the shielded and un-shielded population statistics. As the dispersal of the cloud proceeds further, shielding becomes less effective  and the shielded and un-shielded disc models return to having more similar properties. This is illustrated in the lower panels of Figure \ref{fig:cluster} where the fraction of sinks in the high emission measure (i.e. ionised) parts of the cluster increases over time. 

\begin{figure*}
    \vspace{-0.4cm}
    \hspace{-0.6cm}
     \includegraphics[width=2.0\columnwidth]{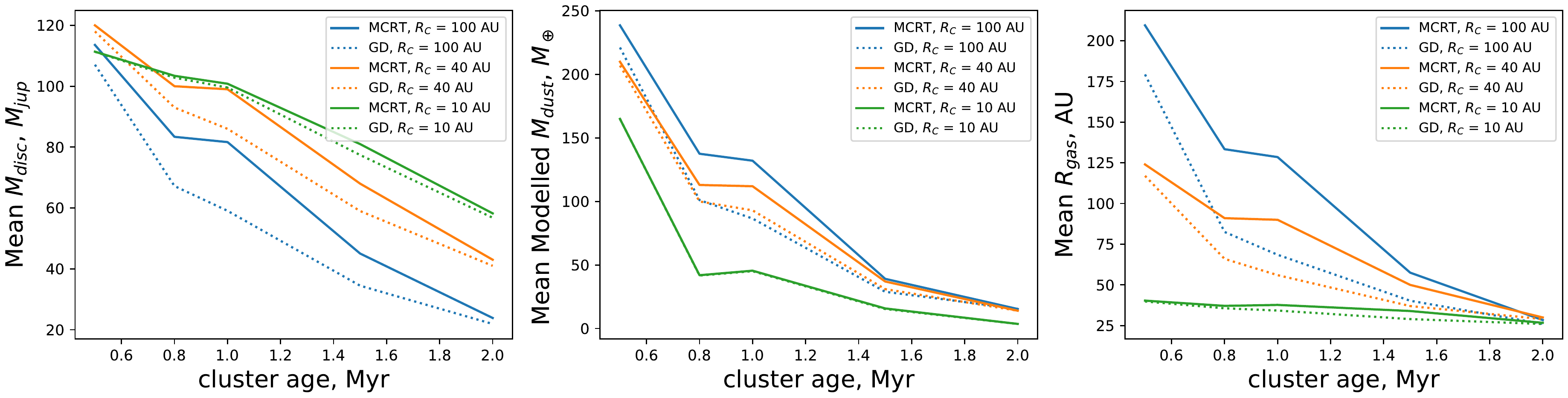}
    \vspace{-0.1cm} 
     \caption{The mean values of disc properties in our cluster population as a function of cluster age. The panels are the gas mass, dust mass and gas disc radius from left to right. }
     \label{fig:mean_values}
\end{figure*}

For total disc mass (first column in figure \ref{fig:cdf_plots}), as mentioned above, shielding has a maximum impact at a cluster age of around 1\,Myr, when the strong UV sources are present but the cloud is not widely dispersed and so there is the largest difference between the two populations at that time. The mean total disc mass is $\sim 13$ $M_{\textrm{jup}}$ more massive than the un-shielded case at 1\,Myr (the mean values over time are shown in Figure \ref{fig:mean_values}). At later times the formation of more UV sources and the dispersal of the cloud leads to irradiation and rapid depletion of the previously shielded discs (see also the left columns of figure \ref{fig:FUV_stats}). Near the end of the simulation at 2\,Myr, there is only 2 $M_{\textrm{jup}}$ difference between in the mean values for the shielded and un-shielded sets (though the period of enhanced mass in the shielded discs could be important in the context of early planet formation as we will discuss in section \ref{sec:planets}. 

For the actual dust mass cdf shown by the blue curves in the middle column, both the MCRT and GD sets show an overall significant decrease in values from 0.8\,Myr to 2\,Myr (with mean $M_{dust}$ = 113 and 100\,$M_{\oplus}$ for MCRT and GD at 0.8\,Myr, and $M_{dust}$ = 14 $M_{\oplus}$ for both set at 2\,Myr shown in the second plot of figure \ref{fig:mean_values}). As discussed in section \ref{sec:detailed}, this is caused by large amounts of dust mass being continuously lost through radial drift in the disc evolutionary model used. However, even without including any dust trapping mechanism, a difference of 19 $M_{\oplus}$ in the mean value is still obtained at 1\,Myr between the MCRT and GD set as a result of gas shielding effect. The effect is largest for the discs with intermediate masses 5-30 $M_{\oplus}$ as the more massive discs are too young for photoevaporation to have removed significant mass in the unshielded discs, while the discs in the low mass tail have evolved long enough to lose most of their dust via the radial drift mechanisms. The orange curves in the middle columns show the observers' equivalent dust mass calculated with equation \ref{equn:M_dobs}. These observers' equivalent dust masses are substantially higher (by an order of magnitude or more) than the true dust mass. We investigated the cause of this inconsistency and found it to be due to the assumption of a disc having constant $T_{\rm dust}$ \citep[also raised as an issue by][]{2021MNRAS.501.3502H} and opacity $\kappa_{\nu}$ at all radii in equation \ref{equn:M_dobs}. With a few sample discs we re-calculated $M_{\textrm{d,obs}}$ in equation \ref{equn:M_dobs} with $T_{\rm dust} = T(R)$ and proper opacity values $\kappa_{\nu}(a_{max})$ (as used in equation \ref{equn:fluxIntegral}), and obtained $M_{\textrm{d,obs}}$ values that have $< 10\%$ difference from the modelled dust mass. This verifies the code for calculating the observed dust mass, and confirms that our discs are optically thin enough for applying equation \ref{equn:M_dobs}. We therefore anticipate that disc masses are being overestimated (potentially by a large margin) when using the continuum,  equation \ref{equn:M_dobs} and the assumption of constant 20\,K temperature and dust opacity from \citet{1990AJ.....99..924B}. However the degree of this discrepancy in reality also depends on the true opacity of the underlying dust in the disc \citep{2018ApJ...869L..45B}. The large opacity difference assumed here results partially from the \citet{1990AJ.....99..924B} opacity (used in equation \ref{equn:M_dobs}) being an extrapolation from infrared to the mm, and otherwise from the use of optical constants from \citet{1996MNRAS.282.1321Z} (in calculating the opacity $\kappa_{\nu}(a)$ used in equation \ref{equn:fluxIntegral}). \citet{1996MNRAS.282.1321Z} describe the carbonaceous component of the dust using amorphous carbon grains which produce a high mm opacity. Alternative choices for the carbonaceous components, such as refractory organics \citep{1996ASPC..104..427S} as used by the `DSHARP' (Disk Substructures at High Angular Resolution Project) opacities may result in lower mm opacities \citep{2018ApJ...869L..45B}, closer to the canonical \citet{1990AJ.....99..924B} value. Establishing what carbon composition is correct is beyond the scope of this work, however here we stick to the opacities from \citet{2019MNRAS.486L..63R, 2019MNRAS.486.4829R} based on \citet{1996MNRAS.282.1321Z} which have had success in reproducing disc flux-radius relations, whereas \citet{2022arXiv220201241Z} find that DSHARP opacities cannot produce sufficiently high fluxes without highly effective trapping.

We also included Lupus \citep{2016ApJ...828...46A} and ONC \citep{2018ApJ...860...77E} continuum distributions in the central column of Figure \ref{fig:cdf_plots}. It is not our objective to undertake a dedicated modelling of the ONC or Lupus (our cluster is more representative of Carina) however we can compare the general behaviour (since we do not yet have comparable data of Carina). The age of the ONC can be partly inferred by the estimated age of $\theta^1$C, which is $2.5\pm 0.5$\,Myr \citep{2005astro.ph.10288S}, and the age of Lupus is $\sim 1-2$ Myr \citep{2008hsf2.book.....R}, but maybe as old as $3\pm 2$ Myr \citep{2014A&A...561A...2A}. The form of the distribution is similar in our models and the observed regions especially for ONC, however the ``observers' equivalent dust masses'' in our models are consistently higher than the observed distributions even at 2\,Myr. These higher observed dust masses hold irrespective of the initial disc size (see appendix \ref{sec:Appendix}) though the difference is smaller for $R_c=10$\,au. This offset is decreasing over time and since the age of $\theta^1$C in ONC $\sim 2.5\pm 0.5$\,Myr, if the cluster simulation were advanced for longer they may become more consistent. Additionally, our sample include all stars/discs in the cluster (including those that would be more massive class 0/I discs), whereas the ONC distribution is based on class II discs only.

The right hand column of Figure \ref{fig:cdf_plots} show the CDF of disc radii. The gas disc radius $R_\textrm{gas}$ (magenta curves) is that which encapsulates 95\,per cent of the mass in the disc. The dust disc radius $R_{68}$ (brown curves) is the radius that contains 68\,per cent of the 850\,$\mu$m emission \citep[e.g. as][]{2021MNRAS.506.2804T}. Note that the distribution of the dust is a function of the grain size and hence continuum wavelength \citep[e.g.][]{2021MNRAS.506.2804T}. The bulk of the dust mass is actually in small grains that are well coupled to the gas (due to quick radial migration of large grains) and so a metric based on the contained dust mass would be similar to that of the gas radius. 

First for the gas radius, at 0.8 and 1\,Myr, the MCRT curves show a large number of discs with large radii due to the newly generated discs in the cluster being shielded from radiation by the gas. Conversely when the shielding is not accounted for the discs are much smaller. The difference between the mean values of the MCRT and GD curves again becomes small as the cluster ages, decreasing from 90 and 56\,au at 0.8\,Myr to 30 and 29\,au at 2\,Myr (shown in the third plot of figure \ref{fig:mean_values}). 

The CDF of 850\,$\mu$m flux radius ($R_{68}$) follow similar shapes as the CDF of the gas radius, but with overall offsets to lower values. The gas disc is known to be more extended than the (sub-)millimetre dust disc both for nearby systems \citep[e.g.][]{2018ApJ...859...21A,2019A&A...629A..79T} and discs in the ONC \citep{2020ApJ...894...74B} which is consistent with these models. At 2\,Myr the 850\,$\mu$m disc radius distribution is very similar to that observed in the $\sim2.5\pm0.5$\,Myr $\theta^1$C region of the ONC.

Overall, although we do not attempt to model any particular system our disc populations are qualitatively similar to those in the ONC at a similar age. They also demonstrate the impact of shielding on the disc statistics and how the effect of shielding is coupled tightly to the formation of high UV sources and their dispersal of the star forming cloud. Shielding is most impactful at the point when there are strong UV sources but before the star forming cloud has been widely dispersed, which in this case is important at around 1\,Myr. Generally this timescale where shielding is important will depend on the type and number of UV sources and the nature of the star formation event. For example the virial ratio of giant molecular clouds can limit the effectiveness of feedback in dispersing the gas cloud, which would favour shielding for longer periods \citep{2012MNRAS.424..377D, 2013MNRAS.430..234D}. 

\section{Discussion}

\subsection{Numbers and fraction of proplyds in a Carina like star forming complex}

From our simulation of externally photoevaporating discs in a Carina-like stellar cluster, we can obtain statistics about the occurrence rates of proplyds at a given cluster age, and hence some theoretical implications of how often proplyds are expected to occur in such type of environments. Note that the term proplyd is actually poorly defined and is often applied to both externally photoevaporating discs and photoevaporating globules. In this section we define ``proplyds'' as discs with mass loss rates due to external photoevaporation $\geq$ $10^{8}$ M$_\odot$\,yr$^{-1}$ \citep[based on estimated mass loss rates for proplyds by]{1999AJ....118.2350H, 2016ApJ...826L..15K, 2021MNRAS.501.3502H}. Although the same mass loss rate could give different morphologies in different environments, we chose to keep this mass loss rate because the difference between the well-observed regions and Carina-like regions is hard to quantify.

Before counting the proplyd frequency, we first calculated the ``disc fraction'' as a function of cluster age, according to the definition in \citet{2019MNRAS.490.5678C} which considers a disc as dispersed if it has lost 90\% of its initial mass. We found that in our model, the disc fraction remained constantly high and is still around 90\% by the cluster age of 2 Myr, mainly due to ongoing star formation in the cluster. However observationally it is quite hard to measure ``disc fraction'', and the disc fraction measurements \citep{2014A&A...561A..54R, 2018MNRAS.477.5191R} are of inner disc fractions which are more likely to be determined by internal winds. Due to the above reasons, we decided when counting proplyds to include discs around all young stellar objects (with with mass $<1.9\,$M$_\odot$) regardless of whether a disc is considered dispersed. 

\begin{figure}
    \centering
	\includegraphics[width=\columnwidth]{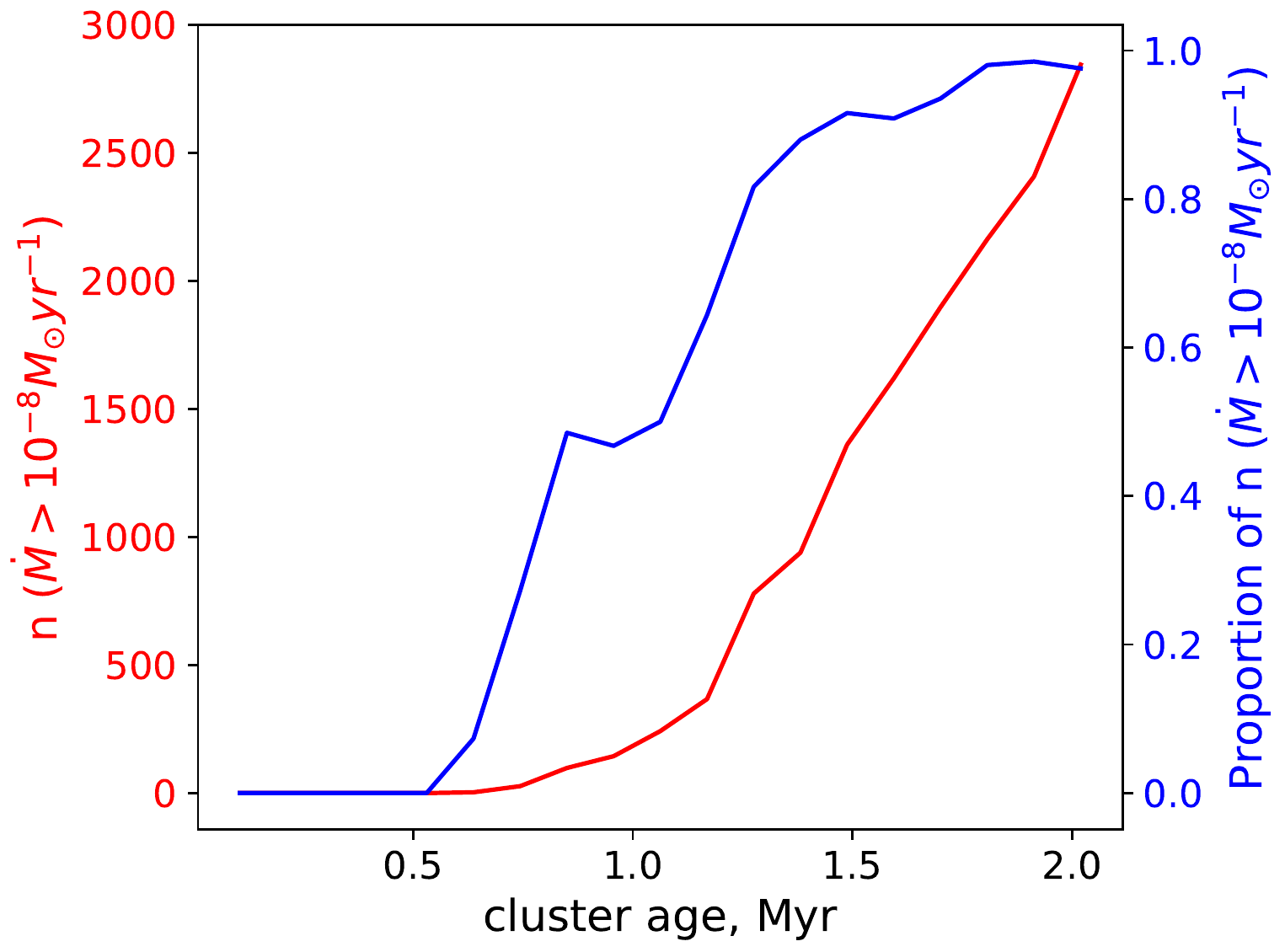}
    \caption{The red curve shows the number of discs with mass loss rates due to external photoevaporation $\geq$ $10^{8}$ M$_\odot$\,yr$^{-1}$ as a function of cluster age, the blue curve shows the proportion of disc with mass loss rates due to external photoevaporation $\geq$ $10^{8} $M$_\odot$\,yr$^{-1}$ as a function of cluster age, for the cluster simulated with initial $R_{C} = 40$\,au, in the Monte Carlo Radiative Transfer case.} 
    \label{fig:num_proplyd}
\end{figure}

The red curve in figure \ref{fig:num_proplyd} shows the number of proplyds as a function of cluster age, which exhibits a rapid increase after 1\,Myr caused by the increased star formation rate (see figure \ref{fig:star_form_rate}) and increasingly widespread dispersal of the star forming gas. The blue curve is the fraction of proplyd discs out of all simulated young stellar objects (with mass $<1.9\,$M$_\odot$) that exist at the specific cluster ages. This also shows an overall fast increasing trend after 0.5\,Myr (caused by generation of a few high FUV sources in the cluster), and suggests by the end of the simulation at 2\,Myr, almost all discs in the cluster are, according to our definition, proplyds. Though note that if the simulation were run for longer than 2\,Myr, the complete dispersal or severe truncation of discs would mean that their mass loss rates drop and the ``proplyd fraction'' would be expected to turn over and start decreasing again. Another caveat is that our stars are represented by cluster sink particles, whereas if these were replaced by a larger number of individual stars a larger fraction may dynamically spread into shielded or weaker UV regions \citep[e.g.][]{2021MNRAS.tmp.3281L}. The offset in time between the rise of the blue and red curves in Figure \ref{fig:star_form_rate} is due to the offset in time at which feedback becomes significant ($\sim0.5$\,Myr) and at which the star formation rate is high ($\sim1$\,Myr).

The two curves in figure \ref{fig:num_proplyd} suggest both a high number and a high fraction of strongly evaporating discs by the age of 2\,Myr in a Carina-like cluster with highly irradiated regions, due to high star formation rates and relatively fast gas dispersal in the cluster (see the discussion in section \ref{sec:UVtracks} about the left panels in figure \ref{fig:FUV_stats}, which illustrates by 2\,Myr in the simulated cluster, most stars are exposed to high FUV radiation field strengths.) The implications of this if true would be significant. Massive star forming complexes form a substantial fraction of the total stellar population, especially at the peak epoch of star formation history, making such complexes representative of the most typical kind of environment \citep[e.g.][]{2014ARA&A..52..415M}. The prediction here is that the majority of discs forming in such complexes (at least on the timescales considered here) are subject to significant external photoevaporation, at a level expected to affect the disc evolution and hence potentially planet formation. 


Our models suggest that a Carina-like environment hosts a large number of strongly evaporating discs. However it will be challenging to observationally verify this prediction since current facilities are not able to resolve proplyds at the distance of Carina \citep[2.3\,kpc][]{2006ApJ...644.1151S}. Cometary objects have been detected towards Carina \citep[][]{2003ApJ...587L.105S, 2010MNRAS.406..952S} and some of these even contain verified embedded protoplanetary discs \citep{2016ApJ...825L..16M}. However those evaporating objects are much larger scale than discs (or proplyds resulting from direct disc external photoevaporation) and so are more likely to be evaporating globules. JWST is sensitive enough to detect circumstellar material out to the magellanic clouds\footnote{There is a JWST GO program to do so \url{https://www.stsci.edu/jwst/phase2-public/1759.pdf}}. However it is unclear whether the brightness of Carina will lead to saturation, nor whether the cluster density or complexities of background subtraction would render observations of discs there infeasible. Even if they are plausible, work needs to be done to determine observational signatures of spatially unresolved propylds at large distances for JWST. There may also be other spatially unresolved means of detecting proplyds at large distances through the strength of emission lines (Ballabio \& Haworth in prep) or by measuring the heating of dust discs \citep[e.g.][]{2005ApJ...631.1134A, 2007ApJ...671.1800A, 2021MNRAS.503.4172H}. Achieving this is key for testing these predictions and understanding how widespread the impact of external photoevaporation truly is.

\subsection{Implications for planet formation}
\label{sec:planets}
Planet formation is a sufficiently complex process that we cannot comment in detail what the impact of these simulations is for planetary populations. However, we can comment broadly on the implications of widespread external photoevaporation and on the role of shielding with the kind of frequency and timescales that our model permits. 

The first point is that terrestrial planet formation around low mass stars \citep[e.g. Trappist-1][]{2017Natur.542..456G} has to be very efficient in order to have sufficient mass in solids to produce the mass in planets that we observe \citep{2017A&A...604A...1O, 2018MNRAS.475.5460H}. Terrestrial planets are found on orbits at scales much smaller than that at which external photoevaporation directly operates, but their formation probably relies on the growth and radial drift of dust from larger radii in the disc since the in-situ mass is not sufficiently high to produce the observed planets \citep{2017A&A...604A...1O}. Dust is removed from the outer disc by external photoevaporation until it grows to sizes where it is no longer entrained \citep[$\sim1-10\mu$m][]{2016MNRAS.457.3593F}. The overall solids mass budget is therefore potentially constrained by external photoevaporation, which according to these models happens in a widespread manner in a Carina-like environment. On the other hand, our simulations do show that shielding can take place for some discs for around 0.5\,Myr, which is sufficient to enable grain growth, drift and potentially the onset of planet formation \citep{2012A&A...539A.148B, 2020Natur.586..228S}, though this becomes less effective for stars formed later in the cluster. Furthermore, \citep{2021arXiv211111077G} observed a $\sim5$\,Myr star with a possible planet embedded in its disc that has just recently emerged from a gas pillar to start being externally irradiated. So some fraction (albeit a small one, we predict) will certainly be embedded for relatively long timescales.

The second broad implication for planet formation is for giant planets forming at larger radii. By warming \citep{2018MNRAS.474..886N}, stripping, and truncating the disc the mass reservoir to be accreted for massive planetary atmospheres is reduced and the migration of planets through the disc could be altered or even halted completely. Again, in a massive star forming complex such as Carina we expect that this could have an impact for the majority of star/discs.  Addressing the impact of external photoevaporation on planet formation in more detail will require dedicated planet formation simulations.

\section{Summary and conclusions}
We performed the first evolutionary models of discs in clusters that are coupled to star cluster formation and photoionisation/radiation pressure feedback simulations. The star formation and feedback calculation is taken from \cite{2021MNRAS.501.4136A}, with size and UV sources representative of a Carina-like star forming region. We track the FUV radiation field at sink particles in the simulation over time using both a simple geometric dilution of the radiation field from other sinks (which ignores shielding due to the cloud) and by using full polychromatic Monte Carlo radiative transfer (which accounts for shielding due to the cloud). These time varying FUV tracks for each star in the model are then coupled with 1D protoplanetary disc viscous evolution and external photoevaporation models. From this we aim to capture the interplay of ongoing star formation, stellar feedback and external photoevaporation on disc populations in massive star forming regions.  We draw the following main conclusions from this work 

1) Even in our calculation that includes gas shielding, most discs are exposed to strong FUV radiation by the time they are 0.5\,Myr old. Embedding any given disc therefore only offers protection from external photo-evaporation for a short time in a massive star forming complex like Carina. 

2) The distribution of FUV fields that discs are exposed to over time is very different in terms of the age of the stellar cluster compared to the ages of individual stars. For example, when the stellar cluster is 1\,Myr old there is a substantial (but young) embedded population that is shielded from the FUV, whereas the vast majority of stars are in a high FUV environment by the time they are 0.5\,Myr old. It is therefore important to distinguish between the cluster age and age of any given disc.  

3) Despite its short duration, shielding by the cloud  prevents early gas and dust loss to external photoevaporative winds and maintains a larger disc radius during the early stage of disc evolution. This could have potentially significant impact if planet formation happens early on, as there is growing evidence for \citep[e.g.][]{2018ApJ...857...18S, 2020Natur.586..228S}. 

4) Shielding is most impactful during the period in which strong UV sources have formed, but the molecular cloud is not yet widely dispersed. In our simulation this is at $\sim1\,$Myr and spans a period of $\sim0.5\,$Myr. During this period shielded discs remain much more extended and retain much more mass than discs that would otherwise be exposed to the FUV radiation field in the region. This could be sufficient to protect from external photo-evaporation and facilitate early planet formation. 

5) From our simulation we expect a high number and a high fraction of strongly evaporating discs by the age of 2 Myr in a highly irradiated region like Carina, which likely represents a typical star-forming environment. Therefore, our results predict that most discs in such regions are subject to external photoevaporation strongly enough to affect disc evolution and planet formation (at least during the first 2 Myr of the cluster age). However it is challenging to observationally confirm this prediction, since this will require methods for detecting external photoevaporation at large distances where propylds are not spatially resolved. 
 
6) Dust grain growth and drift happens very quickly in discs and only small grains are entrained in an external photoevaporative wind. The early shielding hence plays a significant role in preserving the solids mass budget against external photoevaporation. This may affect terrestrial planet formation which replies on growth and radial drift of dust from larger radius in a disc.  However, stripping of gas mass from discs continues even once the dust is too large to be entrained, so the available gas reservoir for planetary migration and atmosphere accretion is predicted to be widely impacted in massive stellar clusters

\section*{Data Availability}
The disc evolutionary code that we used is available on github (\url{https://github.com/AndrewSellek/DiscEvolution}). Simulation data will be made available upon reasonable request.

\section*{Acknowledgements}

TJH is funded by a Royal Society Dorothy Hodgkin Fellowship. AAA acknowledges funding from the European Research Council for the Horizon 2020 ERC consolidator grant project ICYBOB, grant number 818940. ADS thanks the Science and Technology Facilities Council (STFC) for a Ph.D. studentship.

This research utilised Queen Mary's Apocrita HPC facility, supported by QMUL Research-IT (\url{http://doi.org/10.5281/zenodo.438045}).

This research has made use of the NASA Exoplanet Archive, which is operated by the California Institute of Technology, under contract with the National Aeronautics and Space Administration under the Exoplanet Exploration Program.



\bibliographystyle{mnras}
\bibliography{Lin_project} 




\appendix
\section{Statistics of $R_c=10\,$au and $R_c=100\,$au discs}
\label{sec:Appendix}
\begin{figure*}
    \vspace{-0.3cm}
    \hspace{-0.6cm}    
    \includegraphics[width=2.1\columnwidth]{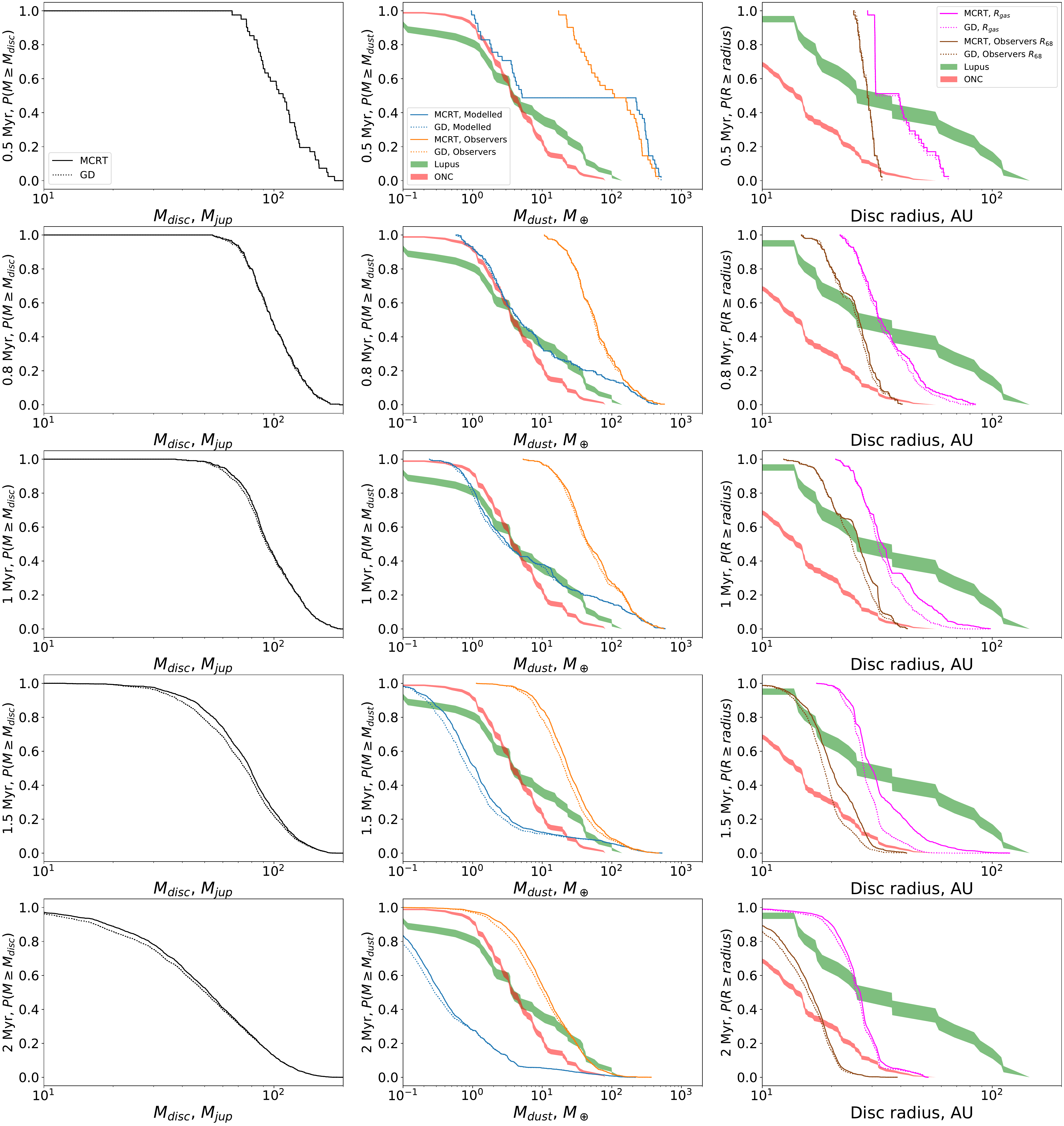}
    \caption{The inverse cumulative distribution functions (cdf) of various disc properties for all discs simulated with initial $R_C$ = 10\,au at 5 cluster ages : 0.5\,Myr (first row), 0.8\,Myr (second row), 1\,Myr (third row), 1.5\,Myr (fourth row) and 2\,Myr (fifth row). The solid curves show the simulated data set with the MCRT tracks and the dotted curves show data set with GD tracks. The first column shows the cdf of the modelled disc mass. The second column shows the modelled dust mass (blue curves), the calculated observes' equivalent dust mass (orange curves), the green and red curves show the observed dust mass distributions for discs in Lupus and ONC. The third column shows the modelled disc gas radius $R_{gas}$ (magenta curves) and observer's equivalent disc radius $R_{68}$ (brown curves), the green and red curves show the observed disc size distributions for discs in Lupus and ONC.}
    \label{fig:cdf_plots_size10}
\end{figure*}

Here we include our population statistics for discs with initial scaling radius of $R_c=10\,$au (Figure \ref{fig:cdf_plots_size10}) and $R_c=100\,$au. Our canonical scaling radius that is the focus of the paper is $R_c=40$\,au (Figure \ref{fig:cdf_plots}).

\begin{figure*}
    \vspace{-0.3cm}
    \hspace{-0.6cm}   
    \includegraphics[width=2.1\columnwidth]{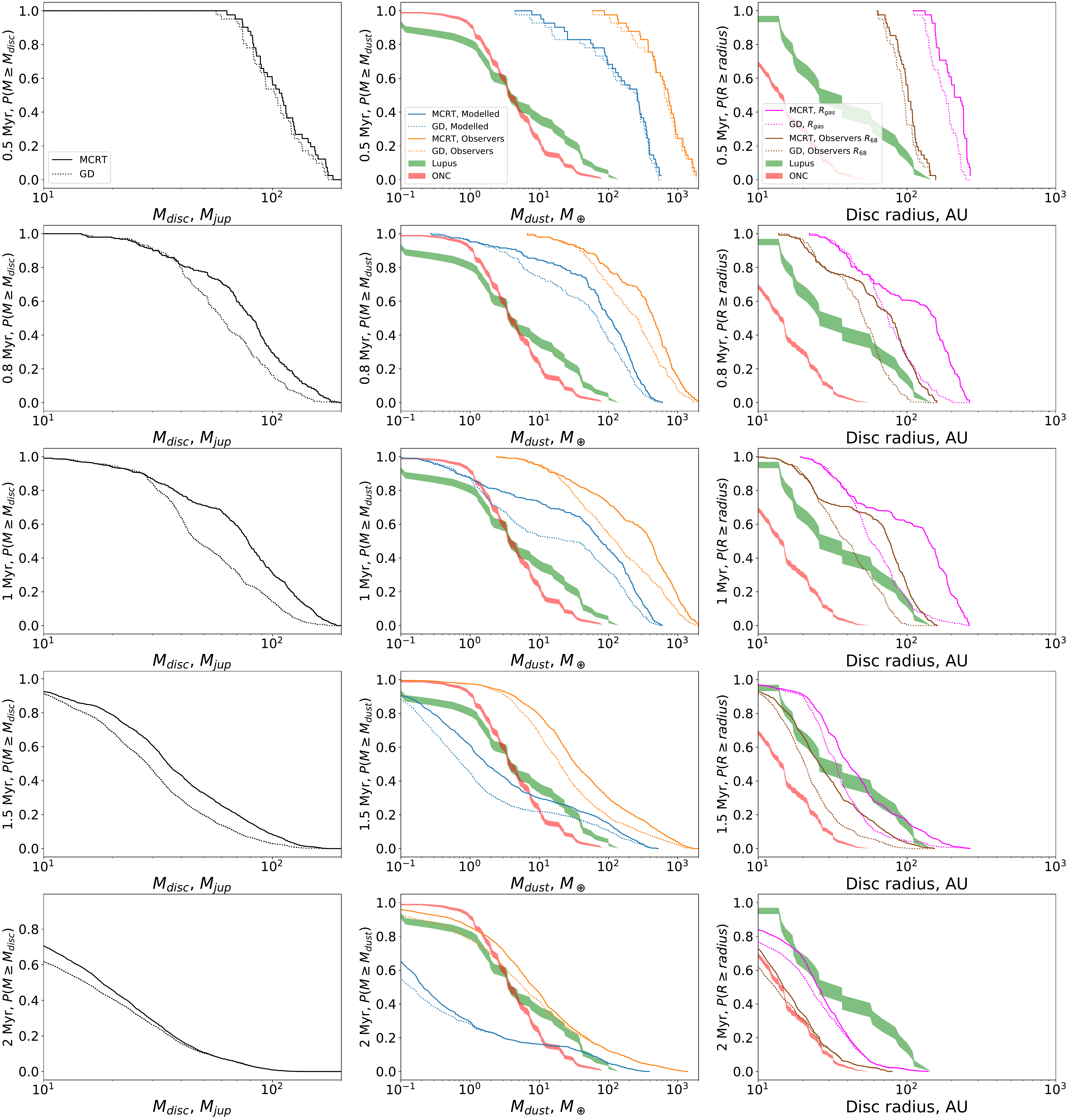}
    \caption{The inverse cumulative distribution functions (cdf) of various disc properties for all discs simulated with initial $R_C$ = 100\,au at 5 cluster ages : 0.5\,Myr (first row), 0.8\,Myr (second row), 1\,Myr (third row), 1.5\,Myr (fourth row) and 2\,Myr (fifth row). The solid curves show the simulated data set with the MCRT tracks and the dotted curves show data set with GD tracks. The first column shows the cdf of the modelled disc mass. The second column shows the modelled dust mass (blue curves), the calculated observes' equivalent dust mass (orange curves), the green and red curves show the observed dust mass distributions for discs in Lupus and ONC. The third column shows the modelled disc gas radius $R_{gas}$ (magenta curves) and observer's equivalent disc radius $R_{68}$ (brown curves), the green and red curves show the observed disc size distributions for discs in Lupus and ONC.}
    \label{fig:cdf_plots_size100}
\end{figure*}


\bsp	
\label{lastpage}
\end{document}